\documentclass[]{aastex631}
\usepackage{graphicx}
\graphicspath{{./}{figures/}}

\newcommand{\cloudy}{{\fontfamily{qcr}\selectfont CLOUDY}}

\begin{document}
\title{New theoretical Fe II templates for bright quasars}
\correspondingauthor{Ashwani Pandey}
\email{ashwanitapan@gmail.com}

\author[0000-0003-3820-0887]{Ashwani Pandey}\thanks{PIFI visiting scientist}
\affiliation{Center for Theoretical Physics, Polish Academy of Sciences, Al. Lotnik\'ow 32/46, 02-668 Warsaw, Poland} 
\affiliation{Key Laboratory for Particle Astrophysics, Institute of High Energy Physics, Chinese Academy of Sciences, 19B Yuquan Road, Beijing 100049, P. R. China}

\author[0000-0002-7843-7689]{Mary Loli Mart\'inez-Aldama}
\affiliation{Astronomy Department, Universidad de Concepci\'on, Barrio Universitario S/N, Concepci\'on 4030000, Chile}
\affiliation{Millennium Institute of
Astrophysics MAS, Nuncio Monsenor Sotero Sanz 100, Of. 104, Providencia, Santiago,Chile}
\affiliation{Millennium Nucleus on Transversal Research and Technology to Explore Supermassive Black Holes (TITANS)}

\author[0000-0001-5848-4333]{Bo\.zena Czerny}
\affiliation{Center for Theoretical Physics, Polish Academy of Sciences, Al. Lotnik\'ow 32/46, 02-668 Warsaw, Poland} 

\author[0000-0002-5854-7426]{Swayamtrupta Panda}\thanks{Gemini Science Fellow}
\affiliation{International Gemini Observatory/NSF NOIRLab, Casilla 603, La Serena, Chile}
\affiliation{Laboratório Nacional de Astrofísica, MCTI, Rua dos Estados Unidos, 154,  37504-364 Itajubá, MG, Brazil}

\author[0000-0001-6450-1187]{Michal Zaja\v{c}ek}
\affiliation{Department of Theoretical Physics and Astrophysics, Faculty of Science, Masaryk University, Kotl\'a\v{r}ská 2, 611 37 Brno, Czech Republic}

\author[0000-0001-9449-9268]{Jian-Min Wang}
\affiliation{Key Laboratory for Particle Astrophysics, Institute of High Energy Physics, Chinese Academy of Sciences, 19B Yuquan Road, Beijing 100049, P. R. China}

\author[0000-0001-5841-9179]{Yan-Rong Li}
\affiliation{Key Laboratory for Particle Astrophysics, Institute of High Energy Physics, Chinese Academy of Sciences, 19B Yuquan Road, Beijing 100049, P. R. China}

\author[0000-0002-5830-3544]{Pu Du}
\affiliation{Key Laboratory for Particle Astrophysics, Institute of High Energy Physics, Chinese Academy of Sciences, 19B Yuquan Road, Beijing 100049, P. R. China}


\begin{abstract}
We present a set of new theoretical Fe II templates for bright quasars covering a wavelength range of 1000-10000 \AA\, based on the recent atomic database available in the C23.00 version of the photoionization code \cloudy. We compute a grid of models for a range of incident photon flux, gas density, and multiple microturbulence velocities. We examine the equivalent widths (EWs) and the ratios of Fe II emission over various wavebands and compare them with observations. Our key results are: (1) The flux generated from the shielded side of the cloud is insufficient to describe the measured Fe II emission. (2) Despite using the newest atomic data we still confirm the long-standing problem that the predicted Fe II UV/optical ratio is significantly larger than that observed in the AGN spectra. (3) The Fe II UV/optical ratio is not significantly affected by the variations in the microturbulence and the metallicity. (4) The microturbulence can create an additional apparent velocity shift of up to 1000 km/s in the spectra. (5) There is no Fe II template based on a single set of physical parameters that can fit the observed UV to optical Fe II emission spectra. We shortly discuss the most likely effects responsible for the Fe II UV/optical mismatch problem: the assumption of the constant density clouds and the heating mechanism for Fe II emitting clouds.
\end{abstract}


\section{Introduction} 
\label{sec:intro}

The dominant component of the bright active galactic nuclei (AGN) in the optical/UV band is the continuum component coming from the cold optically thick geometrically thin disk \citep[see e.g.][for reviews]{krolik_book1999, 2021bhns.confE...1K, 2023Univ....9..492P, 2023arXiv230615082Z,2025arXiv250119365Z}. Locally this component is well described as a power law although at high quality and/or in broadband data some curvature of the component is visible \citep[e.g.][]{capellupo2015}. Superimposed on this continuum component are numerous strong and broad emission lines like H$\beta$, Mg II, C IV, and Ly $\alpha$, depending on the redshift range covered by the data. These lines are easy to identify in the individual spectra of AGN and quasar composite spectra \citep[e.g.][]{francis1991,vandenBerk2001,telfer2002,scott2004,ren2022}. These lines come from the Broad Line Region (BLR) which is located at a fraction of a parsec from the central supermassive black hole (SMBH). This region reprocesses the bulk of the radiation from the central parts very close to the SMBH.

Apart from these well-resolved strong individual lines, we have a contribution from transitions in weakly ionized heavy elements, predominantly iron. These transitions are so numerous that they form a pseudo-continuum well visible in the near-IR (NIR), optical and UV bands. This pseudo-continuum affects considerably the fitting of the strong individual emission lines from BLR. The issue has been identified a long time ago \citep{sargent1968,collin_souffrin1979,wills1985,wills1980}. The important role of Fe II diagnostics has been recognized by \citet{boroson1992} in their studies of the many AGN parameters using the Principal Component Analysis (PCA) approach. Their study led to the identification of the key role of the Eigenvector 1 (EV1) in the quasar sequence. The ratio, $R_{\rm FeII}$, of the equivalent width (EW) of the Fe II in the 4434-4684 \AA\ wavelength range to the EW of H$\beta$ played a key role in EV1, and later it became the foundation of the optical Quasar Main Sequence plots showing the Full Width at Half Maximum (FWHM) of H$\beta$ vs $R_{\rm FeII}$ \citep{sulentic2000,shen2014,marziani2018,panda2018,panda2019,2024ApJS..272...13P}. 

Since Fe II transitions are so numerous, certain standard templates were developed separately for the optical and for the UV bands which were used for detailed modeling of AGN spectra. Some of the templates were purely observational, i.e. created by selecting the object with relatively narrow lines and subtracting the known lines and continuum. The residual was considered as coming from Fe II \citep[e.g.][]{vestergaard2001,marziani2003,park2022,2021Univ....7..484M}. Other templates were mostly theoretical, based on computing the transition rates \citep[e.g.][]{bruhweiler2008,KDP2015,2021PhDT........22P}. Some templates were a combination of both approaches (\citealt{Tsuzuki2006}, but also partially \citealt{KDP2015}). 

The main issue with the empirical or observational templates is that these templates are limited by the signal-to-noise ratio (SNR) and the wavelength coverage of the spectra. There is not yet a single empirical template that can cover the entire optical to UV bands simultaneously.

Recently, \citet{sarkar2021} incorporated three extensive Fe II atomic data sets to the development version of the photoionization code \cloudy \ \citep{2017RMxAA..53..385F}  and examined the predicted Fe II spectra in the UV and optical regions for different SED shapes. They compared their model-predicted Fe II spectra with the observed UV \citep{vestergaard2001}
and optical \citep{2004A&A...417..515V} Fe II templates of  I Zw 1 Seyfert galaxy to constrain the properties of Fe II emitting gas in the BLR. They concluded that the model predicted Fe II spectra for a BLR cloud having gas density $n_{\rm H} = 10^{11}$ cm$^{-3}$, ionizing flux $\Phi_{\rm H}=10^{20}$ cm$^{-2}$ s$^{-1}$, and microturbulence V$_{turb} = 100$ km s$^{-1}$ was consistent with the observed Fe II templates. Additionally, they investigated the $I$(Fe II)/$I$(Mg II) line intensity ratio and found that the ratio predicted by their best-fitted model with solar abundance matched the observed value.

In this paper, we use the latest version C23.00 of \cloudy \ \citep{cloudy23}, which contains all new Fe II atomic databases, to develop a set of new theoretical Fe II templates that cover a wide wavelength range of 1000-10000 \AA. These new templates will be useful to fit the AGN spectra simultaneously from optical to UV bands. 
We analyze the strength of Fe II emission across various commonly used wavebands in both the UV and optical parts of the spectrum, as well as the ratios of Fe II fluxes between these wavebands. Additionally, we assess the equivalent widths of Fe II blends for both the illuminated and shielded parts of the BLR cloud. To understand the role of microturbulence, we explore how varying microturbulence velocities influence the Fe II spectra, focusing on their effect on different Fe II flux ratios. For a fixed value of microturbulence, we further investigate the impact of changing metallicity on the Fe II spectra. Finally, we try to constrain the physical parameters of the BLR gas producing Fe II emission by comparing our model-predicted Fe II templates with the observed spectra of quasars RM 102 and 1H 0413-4031. While \cite{sarkar2021} tested their model-predicted Fe II spectra exclusively on the narrow-line Seyfert 1 galaxy I Zw 1, our objects are more representative of typical quasars.

The paper is organized as follows: Section \ref{sec:method} outlines our methodology, Section \ref{sec:results} presents the results of our analysis, and Section \ref{sect:discussion} provides a discussion of these findings. Finally, a summary of our work is provided in Section \ref{sec:summary}.

\section{Method}
\label{sec:method}
We use the latest version C23.00 of the photoionization code \cloudy\ \citep{cloudy23}. 
\cloudy \ code has four different Fe II atomic data sets reported by \cite{verner1999}, \cite{bautista2015}, \cite{tayal2018} and \cite{smyth2019}. A comparison of these four data sets is presented in \cite{sarkar2021}. For our model calculations, we use the Fe II data set of \cite{smyth2019} which is the default Fe+ dataset in \cloudy \ C23.00. The \cite{smyth2019} Fe II data set includes 716 levels extended up to 26.4 eV generating 255974 emission lines.  

We computed several grids of photoionization models that span five decades in the hydrogen gas density, 9 $\leq \log n_{\rm H}\, (\rm cm^{-3}) \leq$ 14, and the incident hydrogen-ionizing photon flux, 17 $\leq \log \Phi_{\rm H}\,(\rm cm^{-2} s^{-1}) \leq$ 22, with a step size of 0.25 dex. We adopted a fixed hydrogen column density of 10$^{24}$ cm$^{-2}$ and set the abundance to the default solar abundance, as in \citet{sarkar2021}. 
We use the assumption of a constant total hydrogen gas density of the clouds for our photoionization computations. 

The shape of the incident continuum SED plays a crucial role in predicting the Fe II spectra in AGN \citep[e.g.][]{sarkar2021}. The equivalent widths (EWs) of the Fe II UV and optical multiplets, as well as their ratios, are strongly affected by the shape of the incident SED (see Figure 11 of \cite{sarkar2021}. Since we plan to compare our model-predicted Fe II spectra with the observational data of quasar RM 102, which has an Eddington ratio $\lambda_{Edd}\sim$0.33 (see Section \ref{sec:obs_data}), we adopted the intermediate SED of \citet{Jin2012} which is suitable for sources with $\lambda_{Edd}\sim0.28$ for our \cloudy \ simulations. 

We generated such models with different values of microturbulence velocity. We used four values of microturbulence velocity: 0, 20, 50, and 100 km/s. Such a range for the microturbulence velocity was inspired by previous works \citep{baldwin2004, bruhweiler2008, panda2018, panda2021}. 
We also examine the effect of metallicity on the Fe II emission by varying metallicity as 2, 5, and 10 times the solar metallicity. 

\begin{figure}
    \centering
    \includegraphics[width=18cm, height=10cm]{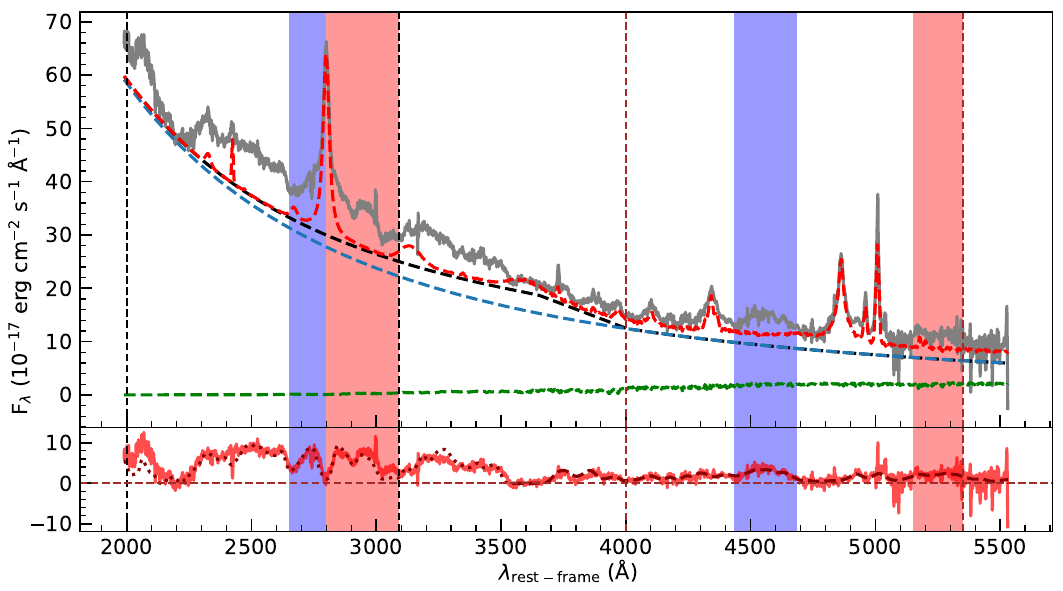}
    \caption{\label{fig:bcz} Observed composite spectrum of RM 102. The regions between the vertical black and brown dashed lines represent the broad UV and optical wavelength ranges, respectively, considered in this work. The shaded blue and red regions denote the blue and red parts in the corresponding regions. Top panel: the observed spectrum is in grey, the dashed red and the blue lines correspond to the best fit of the emission lines and the continuum, respectively. The dashed black line represents the Balmer continuum and the green line denotes the starlight contribution. Bottom panel: the red line depicts the Fe II pseudocontinuum. For reference, we also show the semi-empirical UV (dotted line) and optical (dashed line) templates implemented in \texttt{PyQSOfit} in dark red.
   }
   
\end{figure}

\subsection{Observational data}\label{sec:obs_data}
We provide very preliminary tests of the new templates with the observational data. As for the basic object for these tests, we selected the quasar SDSS~J141352.99+523444.2 (SDSS~RM102) with RA=$14^h13^m53^s$, Dec=$+52^{\circ}34^{\prime}44^{\prime\prime}$, $z$=0.86114$\pm$0.00029, log(L$_{3000}$[erg/s])= 45.0$\pm$0.0005, log(L$_{5100}$[erg/s])= 45.72$\pm$0.0005, log ($M_{BH}$ [M$_{\odot}$])=8.086$\pm$0.036, and an Eddington ratio ($\lambda_{Edd}$) of 0.329$\pm$0.027 \citep{shen2019}.
The spectrum of the source covers a broad wavelength range (2000-5500~\AA). This allowed us to cover both the rest-frame optical Fe II as well as UV Fe II and to test the feasibility of using the same Fe II template to fit the entire UV-to-optical spectral range. 

The size of its BLR region has been estimated by different methods, in this paper we took the one estimated by the Interpolated Cross-Correlation Function (ICCF), $\tau_{{\rm MgII-ICCF}}=101.7^{+11.6} _{+10.3}$ and $\tau_{{\rm H\beta-ICCF}}=104.6^{+20.5} _{+18.5}$ days in the observed-frame \citep{shen2023}. Using the SDSS Science Archive Server (SAS) from the DR16\footnote{\url{https://dr16.sdss.org/optical/spectrum/search}}, we found 70 spectroscopic epochs observed during four years. All the spectra were combined into a median spectrum to obtain a high signal-to-noise spectrum, S/N=30 and 35 at 3000\AA\ and 4000\AA, respectively. Due to the presence of spikes around 5100\AA, it was not possible to estimate the S/N at this wavelength. The median spectrum was corrected for the Galactic reddening using the extinction laws of \citet{cardelli1989} and assuming an extinction value of $A_V = 0.026$ mag in the $V$ band.

\subsubsection{Spectral decomposition of RM 102}\label{sec:disk_fit}
We started with the commonly used approach in which the continuum is modelled as a single power law. Since we could not identify any stellar absorption feature (CaII H and K, G band, Mg~I), we simply assumed that the host galaxy contribution is negligible. 
We considered all the emission lines reported by \citet{vandenBerk2001} between 2000 and 5500~\AA, 31 emission lines in total. We included a model of the Balmer Continuum (BC) plus the High-Order Balmer lines (Bernal et al. in prep.). In the case of the Balmer continuum model, we tested different optical depths ($\tau_{BC}=0.3,$ 1) and electron temperatures ($T_e$=11000, 15000 K) and implemented the Balmer continuum model proposed by \citet{grandi1982}. Regardless of the optical depth and electron temperature considered, the Balmer Continuum contribution was negligible in the spectral fitting and the residuals were consistent in all the cases. Therefore, we decided to use $\tau_{BC}=1$ and $T_e=15\,000$ K, which are the typical values used in large dataset analysis \citep[e.g.][]{kovacevic2014}.  To decrease the number of free parameters, we fixed the same FWHM for each family of emission lines: Balmer lines, Helium transitions, forbidden lines (narrow emission lines), and coronal lines. Flux intensities, FWHM, and shift sum a total of 63 free parameters. We applied Gaussian smearing of FWHM = 4000 km s$^{-1}$ to the UV and optical Fe~II template and a single flux intensity value was taken for the UV and optical wavelengths. In the first attempt, we noticed that it was not possible to fit the UV and optical Fe~II contribution with the same flux intensity. Thus, following the prescription by \citet{boroson1992}, we performed a careful fitting of all the other spectral features and what remained can be considered as the Fe II component. Due to the overlapping of the emission lines and the Fe II pseudocontinuum in some ranges, the flux intensity of the emission line would be underestimated. So, we considered the semiempirical UV/optical Fe~II templates implemented in the fitting code \texttt{PyQSOFit} \citep{pyqsofit} to constrain the intensity of the emission lines. As we previously noticed, even with these empirical Fe II templates, a different flux intensity value is needed to fit the Fe II contribution. The flux intensity factor has a difference of 3.5 between the UV and the optical Fe II contribution. The final spectral fitting is shown in Figure~\ref{fig:MaryLoli} in the appendix. The potential Fe~II contribution is shown at the bottom panel of Figure~\ref{fig:MaryLoli}. As a reference, we also plotted the UV (dotted line) and optical (dashed-line) Fe~II templates implemented in \texttt{PyQSOFit}. 
The optical Fe II shows good agreement, while a slight increase is observed in the UV Fe II within the wavelength ranges of 3000-3150 \AA \ and 3500-3650 \AA. This increase in the UV Fe II could be attributed to the minimal contribution from the BC in this approach. 

We, therefore, performed an alternative analysis by modeling the continuum emission using the standard \citet{SS1973} disk model instead of a power-law. We adopted the black hole mass as given in Section \ref{sec:obs_data} and treated the Eddington rate as a free parameter during the fitting. 
The disk emission exhibits a slightly steeper slope compared to the power law used in the previous approach. To achieve a better fit for the red part of the spectrum, we had to account for some level of contamination from the starlight. 
For the starlight, we used the synthesis stellar model from \citet{vazdekis2016} with a solar metallicity and an age of 7.0795 Gyr appropriate for the host galaxy.  
The normalization of this component was a free parameter. To describe the line emission component, we refitted the line spectrum for a new preliminary shape of the continuum. Later, when fitting the disk and the starlight we did not iterate the line emission, but we allowed for a single parameter which could rescale all the line luminosities while preserving the line profiles and their ratios.

In this approach, we modeled the Balmer continuum following the method of \citet{kovacevic2014}. 
Their method generalizes the model of \citet{grandi1982}, and includes 400 higher-order Balmer lines which considerably contribute to emissions above 3646 \AA. We assumed the temperature of the clouds as $10^4$ K, and the optical depth of the Balmer edge equal to 1. The normalization was a free component of the model.

The new improved spectral fitting of RM 102 obtained from this method is shown in Figure \ref{fig:bcz}, where the expected Fe II contribution from RM 102 closely matches the empirical Fe II templates.

To locate the source in the $\log n_{\rm H}-\log \Phi_{\rm H}$ plane, we introduce several specific spectral windows to measure Fe II emission in all of them separately. The motivation for the band choice is later discussed in Section~\ref{sect:ranges}. These bands are listed in Table~\ref{tab:FeII_blends} and are shown in Figure \ref{fig:bcz}. For comparison, the observed EWs of Fe II in these bands from both spectral fitting methods are provided in Table \ref{tab:FeII_blends}. The EW values obtained using the disk continuum method are slightly lower than those from the power-law method, primarily due to the higher contribution from the BC and the slightly different shape of the continuum. The integrated luminosity of the BC from the second approach is higher than that from the previous decomposition, with the ratio of the BC to H$\beta$ flux equal to $\sim$6.15. We also measured three ratios: red-to-blue wings of UV Fe II, red-to-blue blends of optical Fe II, and  UV to optical Fe II. 
The measured ratios from both spectral fitting are given in Table~\ref{tab:object_Mary_Loli}. 
In our further analysis, we used the values of EWs and ratios from our second (disk continuum) spectral fitting method that provides somewhat better fit to the spectra of RM 102. A comparison of the expected Fe II contribution from RM 102 using both spectral fitting method is presented in Figure \ref{fig:compare_PL_disk_fit} in the appendix.
\subsubsection{{HE 0413-4031}}
\begin{table}
   \centering 
    \caption{\label{tab:FeII_blends}List of Fe II blends used in this work along with their observed equivalent widths for the object RM 102}.   
    \begin{tabular}{|c|c|c|c|} \hline
Fe II blend & Wavelength range (\AA)  &  Observed EW (\AA) &  Observed EW (\AA)\\ 
   &   &   Power-law fit  &   Disk fit \\
\hline
Broadband UV Fe II &	2000-3090  & 290.07 & 266.56\\
UV-blue wing &  2650-2800  & 41.09 & 33.37 \\
UV-red wing &  2800-3090 & 81.22 &  67.50\\
Broadband optical Fe II &	 4000-5350 & 428.82 & 348.36  \\
Optical-blue blend & 4434-4684 & 109.50 & 96.22\\
Optical-red blend &   5150-5350 & 79.27    &   65.04 \\
\hline
\end{tabular}
\end{table}

As a complementary test object, we use the quasar HE 0413-4031. This quasar has been observed with the Southern African Large Telescope (SALT) for 13 years to perform the reverberation mapping of the Mg II line and the UV Fe II \citep{zajacek2020,prince2023,zajacek2023}. In this case, only a very narrow range is covered, between 2700 \AA~ and 2900 \AA~ in the quasar rest frame, and the lines are broader than for RM 102. 

\subsection{Definitions of optical and UV Fe II blends}
\label{sect:ranges}
Since the Fe II emission is of a broadband character the integrated value depends critically on the adopted range of integration. 
We mostly follow the usually adopted values but with some modifications due to the spectral range available for our test object RM 102.

To characterize the broadband emission in the optical range we integrate Fe II from 4000 to 5350 \AA~ \citep{2004A&A...417..515V}. 
Some of the studies concentrate on much narrower ranges. For example, the range from 4434 - 4684 \AA~ was introduced to parameterize the Fe II emission close and bluewards of the H$\beta$ line and to characterize the relative importance of the Fe II to H$\beta$ emission by the corresponding $R_{\rm FeII}$ ratio \citep{boroson1992}. This ratio was well studied observationally in many papers \citep[e.g.][]{osterbrock1977, bergeron1984, shen2014, sniegowska2018, MLMA_etal_2021, marziani2022, 2024ApJS..272...13P, 2024ApJS..272...11P}.

Broad UV Fe II range can be well characterized by the Fe II emission integrated in the wavelength range from 1250 to 3090  \AA \ \citep{vestergaard2001}. We modified this range shortening it on the UV side since the test object did not cover the spectra below 2000 \AA. This range contains the Fe II bump at $\sim$ 2500 \AA \ seen in the spectra as well as in early models of Fe II emission \citep{wills1985}. 

We also conduct a systematic analysis of two UV and two optical wavelength ranges that may be used to determine the spectral shape. The UV ranges, 2650-2800 \AA \ and 2800-3090 \AA \, characterize the Fe II emission on each side of the Mg II line. The optical ranges are 4434-4684 \AA \ and 5150-5350 \AA \, corresponding to the Fe II emission at the blue and red sides of the H$\beta$ line, respectively. The Fe II emission on both sides of Mg II forms clear local bumps or peaks. Therefore, for convenience, we refer to these regions as UV Fe II blue and red wings \citep[e.g.][]{sarkar2021}. The Fe II emission on the blue and red sides of H$\beta$ forms larger blends, which we call optical blue and optical red blends, respectively.
A list of Fe II blends considered in this work is given in Table \ref{tab:FeII_blends}.

\begin{figure*}
\centering
\includegraphics[width=18cm, height=16cm]{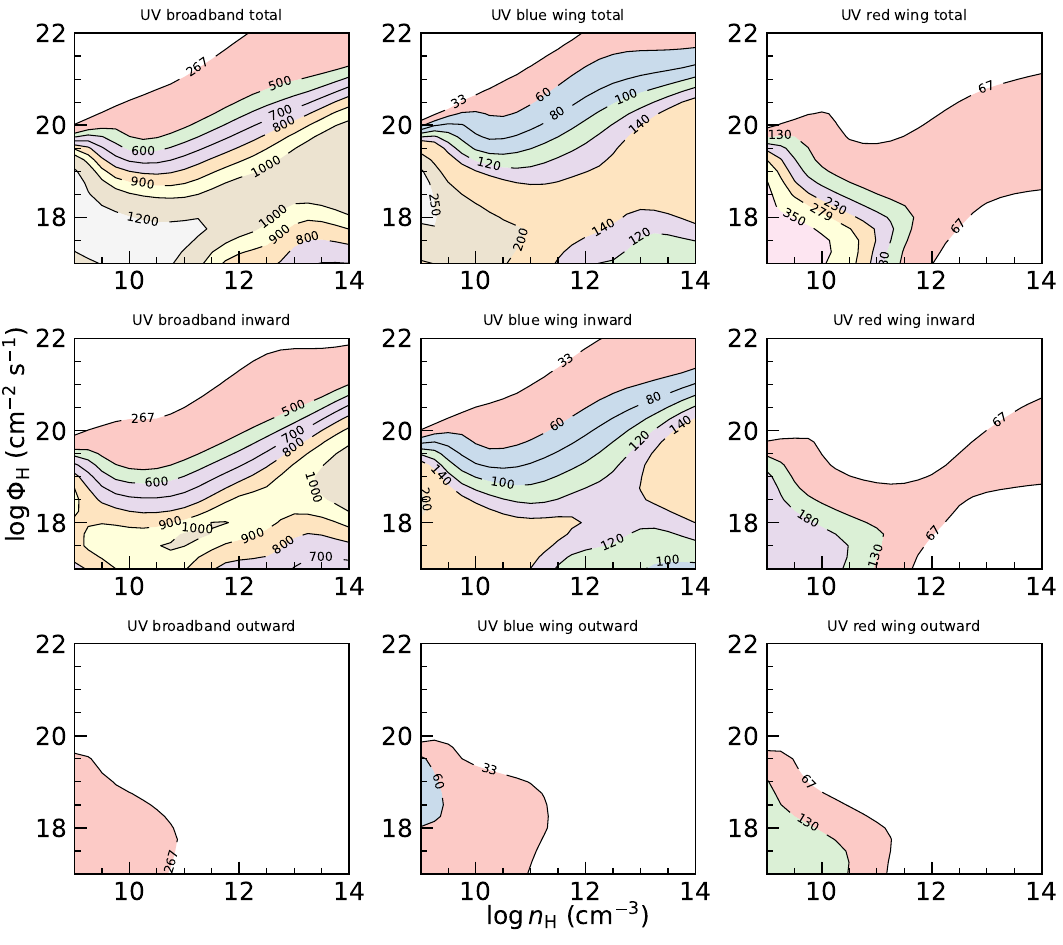}
\caption{\label{fig:EW_uv_turb0}Contour plots for the EW of UV Fe II blends without microturbulence. We only plot contours with EW greater than the observed EW.}
\end{figure*}

\begin{figure*}
\centering
\includegraphics[width=18cm, height=16cm]{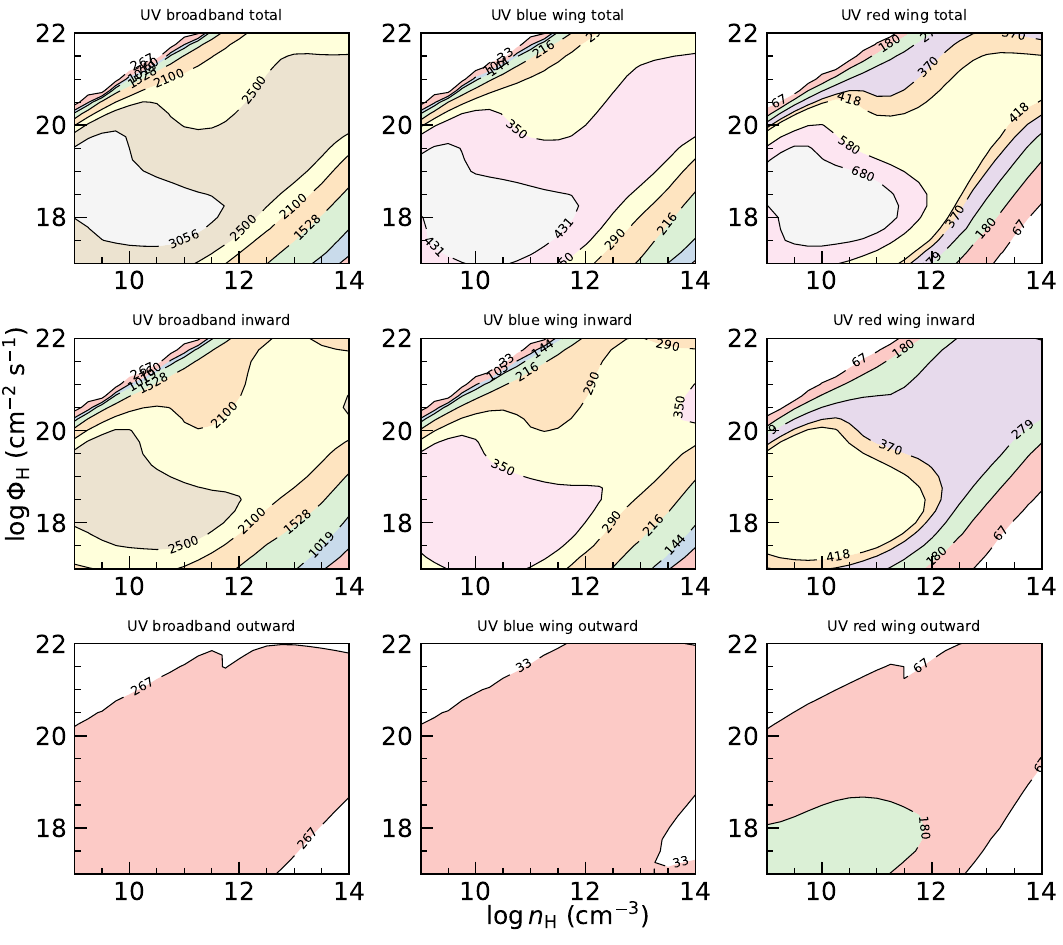}
\caption{\label{fig:EW_uv_turb100}Contour plots for the EW of UV Fe II blends for microturbulence V$_{\rm turb} = 100$ km/s. }
\end{figure*}

\begin{figure*}
\centering
\includegraphics[width=18cm, height=16cm]{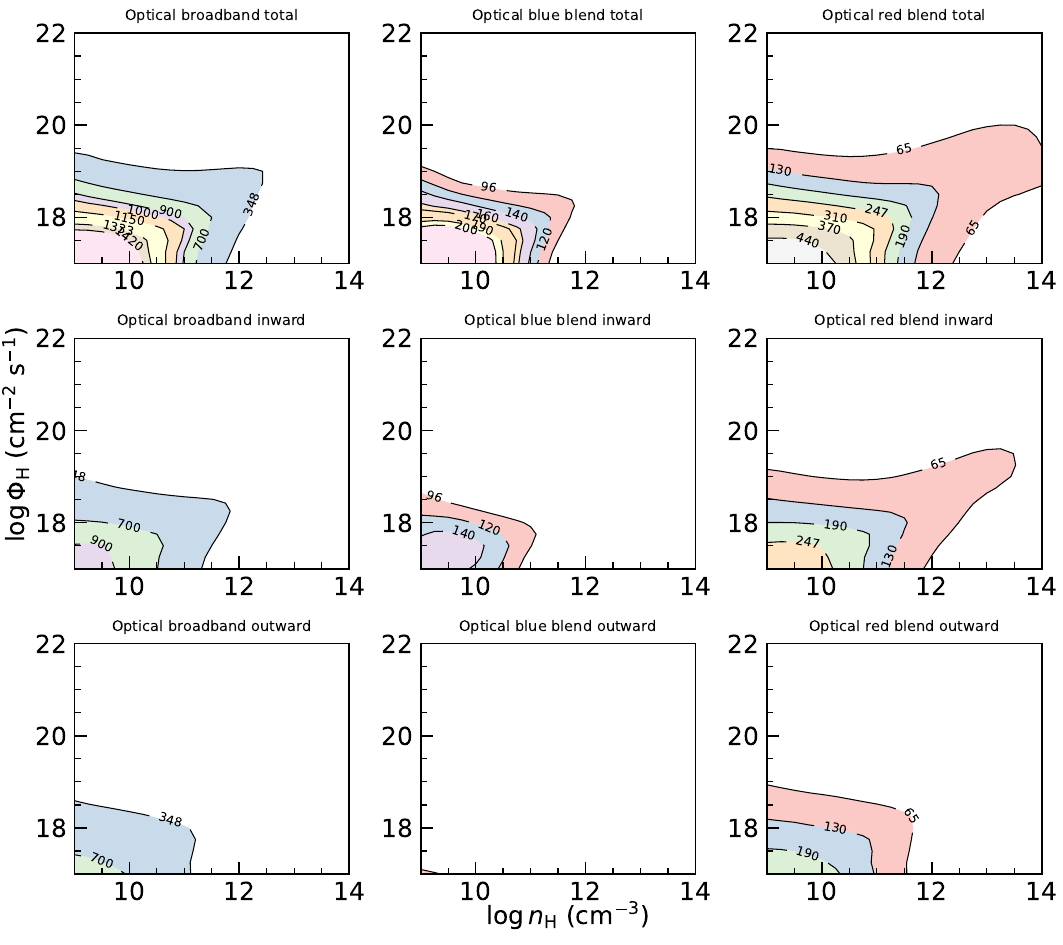}
\caption{\label{fig:EW_op_turb0}Contour plots for the EW of Optical Fe II blends without microturbulence. }
\end{figure*}

\begin{figure*}
\centering
\includegraphics[width=18cm, height=16cm]{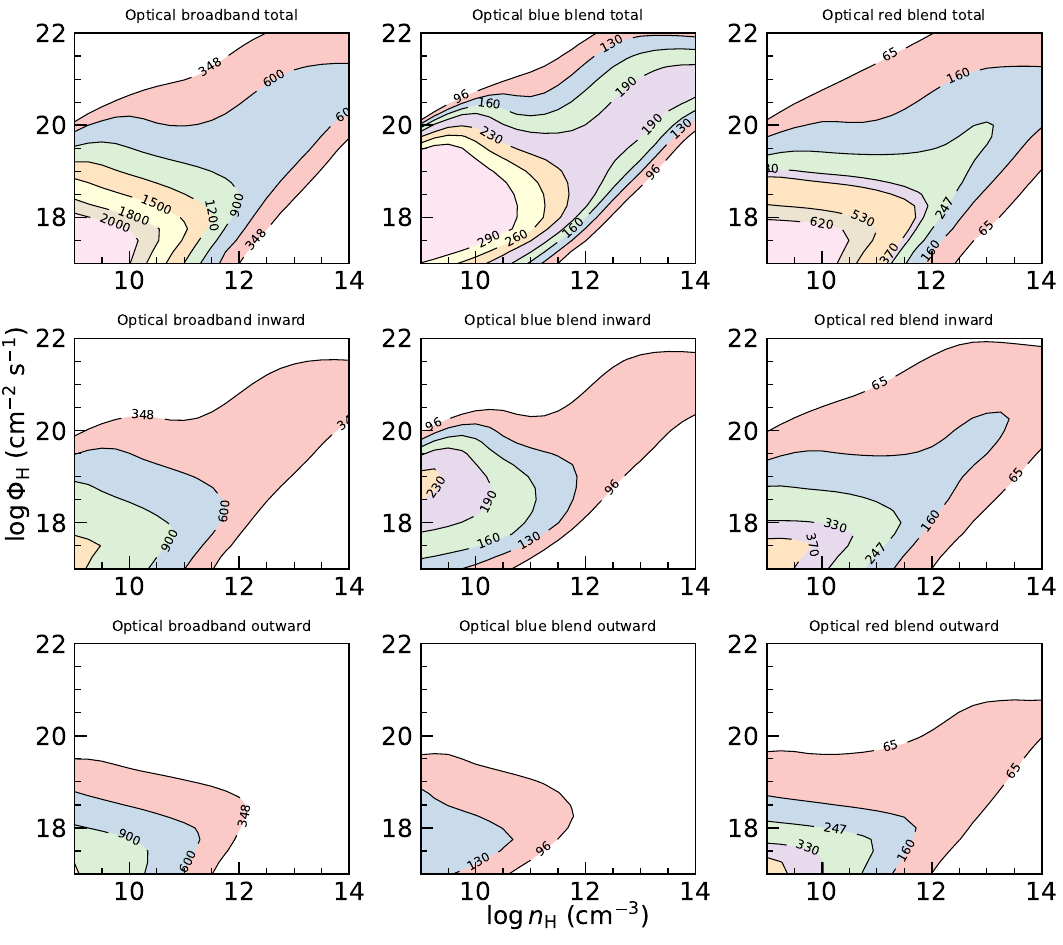}
\caption{\label{fig:EW_op_turb100}Contour plots for the EW of Optical Fe II blends for microturbulence V$_{\rm turb} = 100$ km/s. }
\end{figure*}

\begin{table}
    \centering 
    \caption{\label{tab:object_Mary_Loli} The observed values of different Fe II flux ratios for the test object RM102.}    
    \begin{tabular}{|c|c|c|c|c|} \hline
Ratio & Numerator range (\AA) &  Denominator range (\AA) & Observed value  & Observed value \\
      &                        &                        & Power-law fit    & Disk fit \\
\hline
Spike-to-gap &  (2280–2430) and (2560-2660) & 2430-2560  &  1.65 & 1.58 \\
UV-to-optical & 2000-3090     &  4000-5350               &  2.04 & 2.54 \\
UV red-to-blue wings & 2800-3090 & 2650-2800            &  1.98  & 2.02 \\
Optical red-to-blue blends & 5150-5350       &  4434-4684 & 0.72 &  0.68 \\
\hline
    \end{tabular}
\end{table}

\section{Results}\label{sec:results}
 We begin by examining the strength of Fe II emission, which is affected by the covering factor representing the fraction of the central flux intercepted by the BLR. This limits the parameter space as some parameter combinations do not result in the effective generation of Fe II even for a 100\% covering factor.
Next, we discuss the issue of the difference between the observed and model-predicted spectral shape of the UV Fe II emission. We then discuss the different Fe II  ratios as described in Section~\ref{sect:ranges}. These ratios are independent of the issue of the covering factor. This is followed by the discussion of the ratio of the Fe II to  H$\beta$ which is broadly used to parameterize the quasar population.  Finally, we test the new templates in more detail against our two test objects, showing the exemplary spectral fits.

\subsection{Equivalent widths of Fe II blends}
For our model-predicted spectra, we estimate the equivalent widths (EWs) of all the Fe II blends, given in Table \ref{tab:FeII_blends}, by first computing the integrated Fe II flux in the corresponding wavelength range and then dividing this integrated flux by the incident continuum at a reference wavelength. To minimize the systematic effects associated with the assumed shape of the incident continuum, we choose 3000 \AA \ and 5100 \AA \ as the reference wavelengths in the UV and optical regime, respectively. 
Using a similar process, we also compute the EWs of all the blends for the observed spectra of RM 102 and mention them in Table \ref{tab:FeII_blends}. Since \cite{ferland2009} suggested that the observed Fe II emission could originate from the shielded part of the cloud, in addition to the EW contours of total Fe II emission, we plot EW contours for the inward and outward Fe II emissions in Figures \ref{fig:EW_uv_turb0}-\ref{fig:EW_op_turb100} for microturbulence velocities 0 and 100 km/s. The EW contours for the microturbulence velocities of 20 and 50 km/s are shown in Appendix \ref{appendix:A}. In our model calculations, we assumed that each model cloud within the density and ionizing photon flux grid has a covering factor of 100\% and thereby covers 4$\pi$ steradians of the sky as seen by the (essentially point) source of ionizing photons.  But the covering factor for the BLR cloud has a much smaller value—roughly $\sim$10-30\%  \citep{baldwin1995,2019MNRAS.489.5284K,panda2021}. Thus, the observed EW corresponds to the lower limit of the predicted EW.
Therefore, the region consisting of EW higher than the observed EW should only be consistent with the observations. 

The colored contours in Figures \ref{fig:EW_uv_turb0}-\ref{fig:EW_op_turb100} mark the regions in the n$_{\rm H}$-$\Phi_{\rm H}$ plane whose clouds predict, assuming a 100\% covering factor, at least the measured EWs of the various Fe II blends. The clouds corresponding to the models in the empty regions predict too little Fe II emission line strength in the specific Fe II blend, even for a 100\% covering factor. Moreover, since the BLR emission does not originate from a single set of cloud parameters with a single source covering fraction, the EW contours in Figures \ref{fig:EW_uv_turb0}-\ref{fig:EW_op_turb100} are indicative of the cloud parameters that can significantly contribute to the observed Fe II emission line spectrum. The strength of Fe II emission is enhanced when the microturbulence is high, as seen from the comparison of Figure~\ref{fig:EW_uv_turb0} and Figure~\ref{fig:EW_uv_turb100}. However, even with microturbulence 100 km s$^{-1}$ the outward emission is very weak in the UV.  For the broadband and the blue UV wing even with the assumption of a 50\% covering factor no allowed region appears for outward emission. In the case of the red UV wing, there is a small allowed region (contour marked as EW=180 \AA) for outward emission that corresponds to the covering factor of $\sim$38 \%, but it appears only for low values of the ionizing photon flux. Thus, at least about object RM 102, we can conclude that the shielded sides of the clouds cannot be the dominant source of the Fe II emission in UV.

The strength of integrated Fe II emission in the optical band is overall higher when measured in terms of the EW and lower in terms of absolute flux (see Figures~\ref{fig:EW_op_turb0} and \ref{fig:EW_op_turb100}). Without microturbulence, only the low ionizing photon flux and relatively low-density lead to significant flux. Microturbulence again enhances the Fe II emission strength. The difference between the inward and outward emission is smaller than that for UV. Still, the non-irradiated sides of the clouds are not likely to emit enough radiation to explain the observed Fe II emission when the BLR covering factor of the order of 10 - 30\% is considered. The small allowed region requires the covering of order of 50\% for the total optical Fe II, it is only slightly broader for the red blend and never corresponds to the data from the object RM 102 for the blue blend.

Therefore, in further studies, we concentrate on the total Fe II flux and do not show separate contour plots for the outward radiation coming from the unilluminated cloud side.

\begin{figure*}
\centering
\includegraphics[width=18cm, height=6cm]{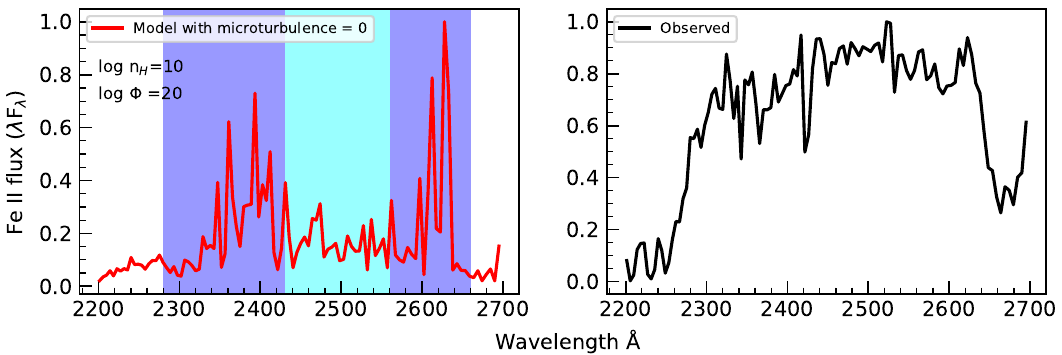}
\caption{\label{fig:spike_spectra}A comparison of observed and model predicted spectra in the wavelength range 2200-2700 \AA. Both spectra are scaled to 1 at their maximum values. (a) Spikes are seen at around 2400 and 2600 \AA \ in the model-predicted spectra without including microturbulence. Blue-shaded regions denote the wavelength range for spikes while cyan-shaded region shows that for the gap. (b) Observed spectrum of RM 102 indicating no such spikes.}
\end{figure*}

\begin{figure*}
\centering
\includegraphics[width=14cm, height=12cm]{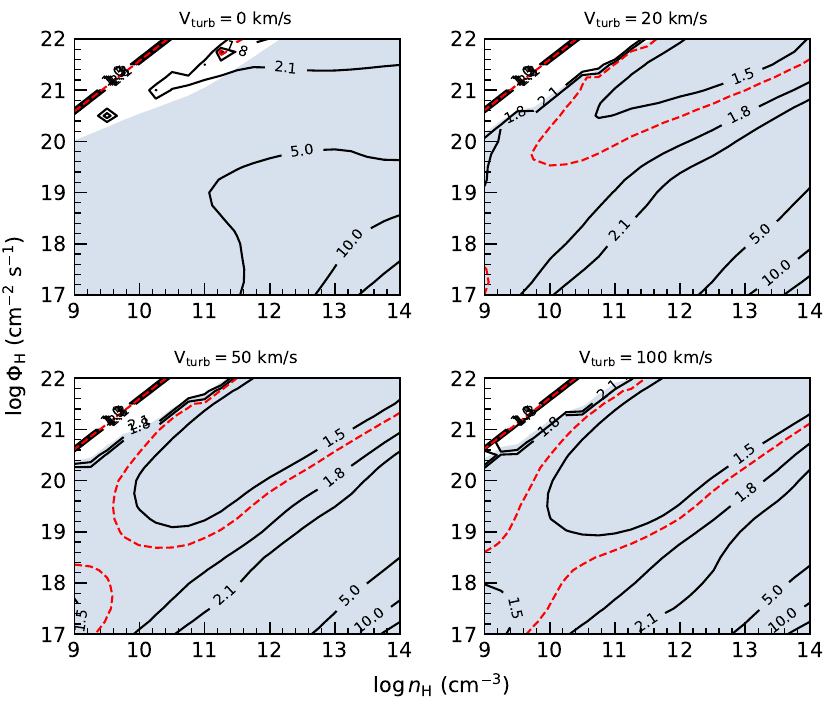}
\caption{\label{fig:spike_gap}Contour plots for spike-to-gap ratio for different microturbulence velocities. The shaded region represents the model parameter space that predicts the EW of broadband UV Fe II consistent with the observed value, assuming a covering factor of 100\%, for the test object RM 102. The red dashed contour denotes the observed value (1.58) of the ratio for source RM 102.}
\end{figure*}

\subsection{Shape of UV Fe II emission}
\cite{baldwin2004} noticed two strong spikes at around 2600 and 2400 \AA, resulting from UV 1 and UV 2+3 multiples, respectively, in the model-predicted spectra that are absent in the observed spectra (see Figure \ref{fig:spike_spectra}).  
In the standard photoionization model (including thermal velocity only), the optical depths at the illuminated surface of the cloud are small. Consequently, the energy deposited by the incident photons is radiated through the strongest resonance lines near 2400 and 2600 \AA \ only that appear as the two spikes in the model spectrum of the cloud.
The presence of these spikes in the model spectra is because Fe II mostly forms in the illuminated face of the cloud with significantly smaller optical depths, which permits the strongest resonance lines to exit directly. This issue of the difference in the shape of Fe II emission was largely investigated by \cite{baldwin2004} and recently, by \cite{sarkar2021}. They suggested that including the microturbulence ($\geq$100 km/s) into photoionization calculations can solve this shape problem. When microturbulence is incorporated into the model cloud, it increases Doppler broadening, which lowers the line optical depths. As a result, lines can escape more easily and appear brighter. Moreover, microturbulence increases the likelihood of continuum and wavelength coincident bound-bound pumping of levels upward (including the upper level of the original transition), since the line can now absorb photons over a larger Doppler width \citep{wills1985,Bottorff2000}.

To examine this UV Fe II shape issue, we adopted the definition of the spike as the integrated Fe II flux over the wavelength ranges 2280–2430 \AA \ and 2560–2660 \AA \ and that of the gap as the integrated Fe II flux over 2430–2560 \AA \ wavelength range from \cite{baldwin2004} and computed spike/gap ratio for all the models. The contour plots for the spike/gap ratio for all the microturbulence velocities are shown in Figure \ref{fig:spike_gap}. 
The observed value of the spike/gap ratio for the spectrum of RM 102 is 1.58 (see Table~\ref{tab:object_Mary_Loli}) and corresponds to the red dashed contours in Figure~\ref{fig:spike_gap}. When the microturbulence is zero, no such contour is seen. However, including microturbulence in our model calculations affects the shape of the UV Fe II and contours that match the measured value of the spike/gap ratio are visible, for a wide range of model parameters. 
So again, higher values of the microturbulence velocities are favoured.

\begin{figure*}
\centering
\includegraphics[width=14cm, height=12cm]{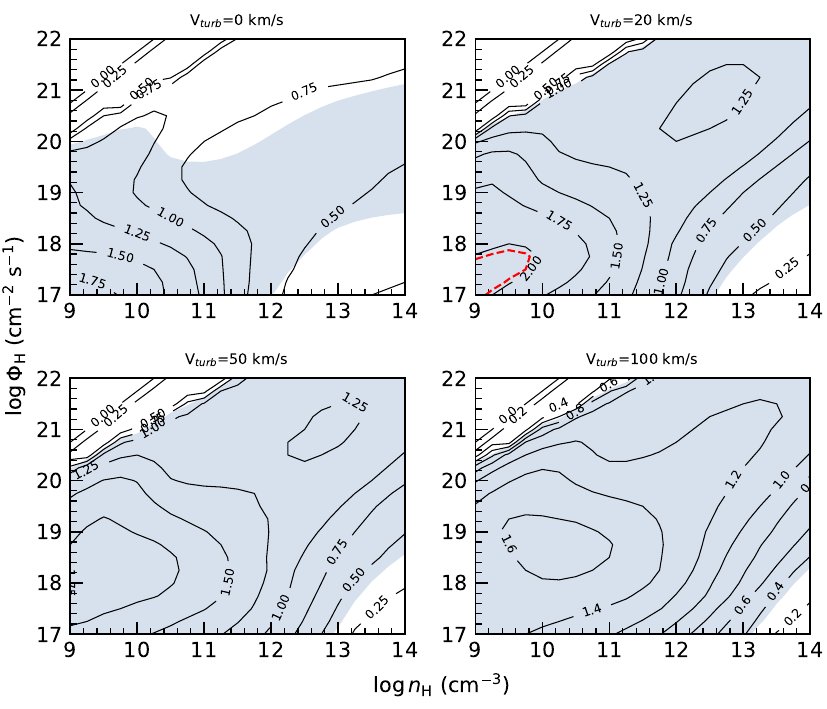}
\caption{\label{fig:red_blue_ratio}Contour plots for the ratio of red (2800-3090 \AA) to blue (2650-2800 \AA) part of the UV Fe II spectra for different microturbulence velocities. The shaded region denotes the model parameter space that predicts the EW of UV red Fe II wing consistent with the observed value for a 100\% covering factor. The red dashed contour denotes the observed value (2.02) for source RM 102.}
\end{figure*}

\begin{figure*}
\centering
\includegraphics[width=18cm, height=5cm]{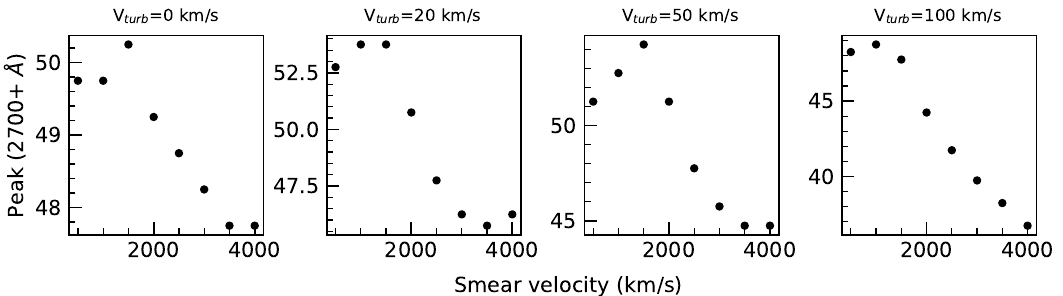}\\
\includegraphics[width=18cm, height=5cm]{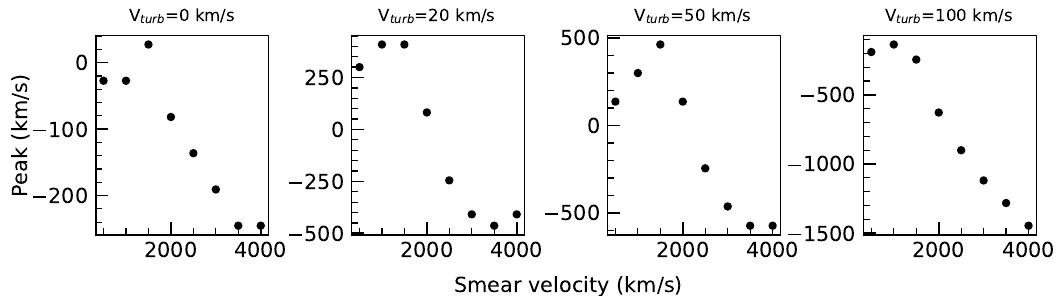}
\caption{\label{fig:peak_width}Variation of UV Fe II peak wavelength with the smear velocity for $\log \Phi_{\rm H}\,(\rm cm^{-2} s^{-1}) = 20.5, \log n_{\rm H}\, (\rm cm^{-3}) = 12$,  and different microturbulence velocities. Top panel: UV peak in \AA. Bottom panel: UV peak in velocity space. Here, a negative peak value represents an outflow, while a positive value denotes an inflow. } 
\end{figure*}

In this study, we primarily examined six Fe II blends (listed in Table \ref{tab:FeII_blends}). The model parameter space in the $\log$n$_{\rm H}$- $\log$$\Phi_{\rm H}$ plane that predicts the EW (for a 100\% covering factor) of Fe II consistent with the observed value differs for each blend. For instance, the parameter space for the optical Fe II broadband blend is more constrained than that for the UV Fe II broadband blend (see Figures \ref{fig:EW_uv_turb100} and \ref{fig:EW_op_turb100}). Therefore, in Figures \ref{fig:spike_gap}–\ref{fig:vary_Z_fe2_hbeta_ratio}, we have shaded only the region corresponding to the Fe II blend with the narrower parameter space among those Fe II blends used to calculate the ratio, except in Figures \ref{fig:spike_gap}, \ref{fig:fe2_hbeta_ratio} and \ref{fig:vary_Z_fe2_hbeta_ratio}. In Figure \ref{fig:spike_gap}, the shaded region is consistent with the Fe II emission strength for the broadband UV Fe II blend, whereas in Figures \ref{fig:fe2_hbeta_ratio} and \ref{fig:vary_Z_fe2_hbeta_ratio} the shaded region specifically denotes the models that predict the observed optical blue Fe II for a covering factor of 100\%.

\subsection{Red-to-blue UV Fe II wings ratio}
In our next plot, we concentrate on the global properties of the UV part of Fe II emission which is important for the proper modeling of the Mg II line. The Fe II emission under the Mg II line is never close to zero, even right under the central parts of the Mg II, at 2800 \AA, as adopted in the creation of UV Fe II observational templates \citep[e.g.][]{vestergaard2001}. However, due to the overlap of the Fe II and Mg II there, it is much easier to measure the strong wings that Fe II forms at both the red and blue sides of Mg II. Proper modeling of the underlying Fe II may affect the conclusions about the dynamics of the Mg II emitting region \citep[e.g.][]{jonic2016,popovic2019}.

The contour plots for the ratio of red (2800-3090 \AA) to blue (2650-2800 \AA) wings of UV Fe II, shown in Figure~\ref{fig:red_blue_ratio}, exhibit a complicated structure. 
At low ionizing photon flux, the ratio depends mostly on the density. At higher ionizing photon flux the dependence is more complex. We also observe the dependence on the microturbulence velocity, as in the central parts the ratio is initially rising, and finally decreasing when the microturbulence velocity crosses the value 50 km/s. 
The measured value of the ratio (shown by the red dashed contour) for the test source is 2.02 which can only be found for a very narrow range of parameters for the microturbulence velocity of 20 km/s.
The two wings represent different multiplets in templates of \citet{KDP2015}, so they may, or may not, come from the same region (see Section~\ref{sect:discussion}). 

\subsection{The effect of the UV Fe II broadening on the measurements of the outflow properties}

In the analysis of their Sloan Digital Sky Survey (SDSS) sample, \citet{KDP2015}  noticed a systematic averaged redshift in the UV Fe II to the optical Fe II. To check if such a shift can be caused just by the shape of the Fe II emission when broadened with the microturbulence effect included, we used a specific Fe II template characterized by the ionizing photon flux, $\log \Phi_{\rm H}\,(\rm cm^{-2} s^{-1}) = 20.5$, and the density, $\log n_{\rm H}\, (\rm cm^{-3}) = 12$. We applied a Gaussian smearing of 2000 km s$^{-1}$ to the template and determined the peak wavelength of the blue wing of UV Fe II. The peak is usually located close to $\sim 2750$ \AA~, and it is dominated by the multiplets 62 and 63 (see \citealt{KDP2015}). However, when we model the entire Fe II emission of the blue wing we see that the position of the peaks moves with the adopted smear velocity. We plot this effect simply in \AA~units (top panel), and in velocity space (bottom panel) in Figure~\ref{fig:peak_width}. The shift corresponds to the apparent velocity from 100 km s$^{-1}$ up to 300 km s$^{-1}$ when the microturbulence is not included, and the presence of the microturbulence affects the apparent direction of the outflow and its amplitude. Apparent velocity shifts up to 1000 km s$^{-1}$ are possible even though no velocity is present in the model which corresponds in each case to the rest-frame spectrum. In \citet{KDP2015} more than half of the sources show a kinematic offset above 1000 km s$^{-1}$. We do not see such large shifts due to the Fe II smoothing but we experimented only with a single template. 

\begin{figure*}
\centering
\includegraphics[width=14cm, height=12cm]{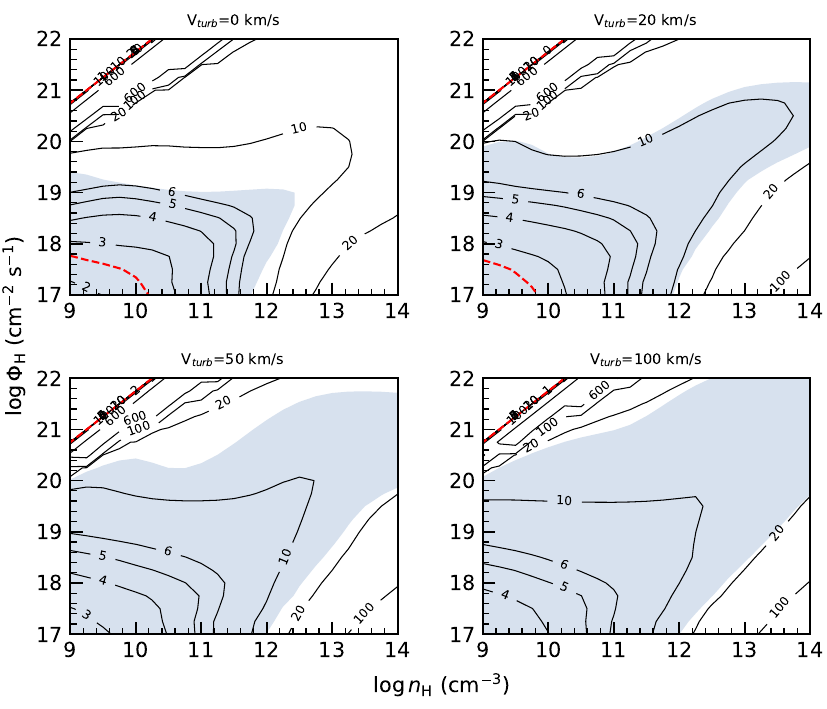}
\caption{\label{fig:uv_optical_fe2_ratio} Contour plots for the ratio of UV to optical broadband Fe II emission for different microturbulence velocities. The shaded region represents the parameter space reproducing the optical broadband Fe II emission strength in test object RM 102, for a covering factor of 100\%. However, the observed value (2.54) of the ratio in this object, shown by the red contour, is barely reproduced in the case of low microturbulence, and it is never consistent with the model for high microturbulence.}
\end{figure*}

\subsection{UV-to-optical Fe II emission}
We calculated the global ratio of the UV to optical broadband Fe II  emission with the integration range specified in Section~\ref{sect:ranges}. The results are shown in Figure~\ref{fig:uv_optical_fe2_ratio} using contour plots. The representative values in the central parts of the plot are in the range 6 to 10.

As already argued by \citet{Joly1981J} and \citet{wills1985},  a ratio of the order of 1 in the relative intensities of the optical to UV lines implies a column density of about $10^{23}$ to $10^{24}$ cm$^{-2}$ to balance Fe II line optical depths and the Balmer opacity. In our computations, we use the value of $10^{24}$ cm$^{-2}$ but we hardly recover the ratio measured for our reference object RM 102. 
We find that this ratio grows rapidly as the ionizing photon flux and cloud density increase. 
\citet{panda2019_WC} and \cite{sarkar2021} stressed the role of the microturbulence velocity in the Fe II intensity but the global ratio of the UV to optical is relatively unaffected. 

The ratio shown in Figure~\ref{fig:uv_optical_fe2_ratio} represents the expectations based on the assumption of a single cloud responsible for both UV and optical emission, which may not be the case in actual objects. We address this issue in the discussion.

\begin{figure*}
\centering
\includegraphics[width=14cm, height=12cm]{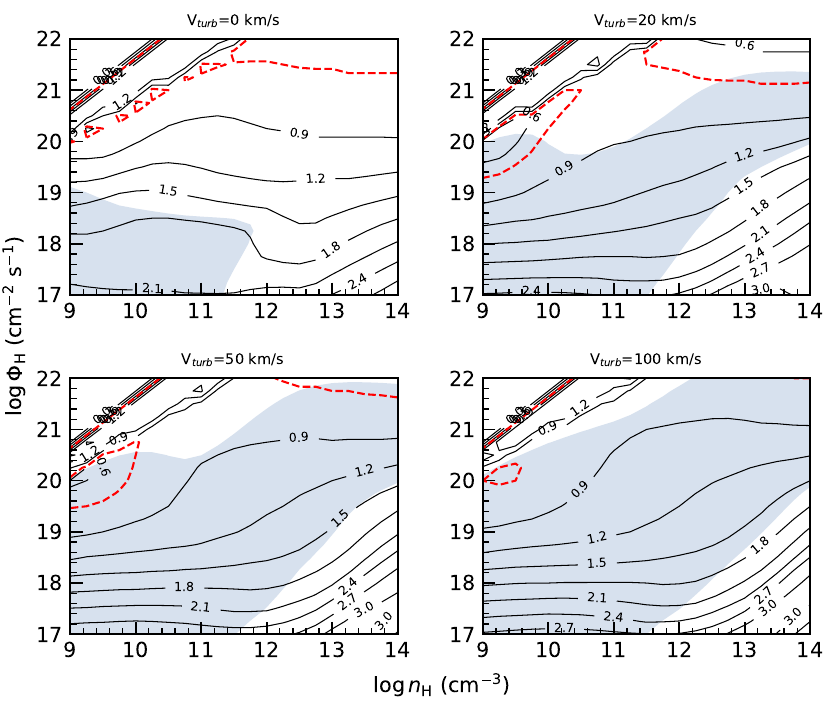}
\caption{\label{fig:red_blue_ratio_hbeta}Contour plots for the ratio of red (5150-5350 \AA) to blue (4434-4684 \AA) part of the optical Fe II spectra for different microturbulence velocities. The shaded region indicates the model parameter space that predicts the strength of optical blue Fe II emission, assuming a covering factor of 100\%, consistent with the observed value. The red dashed contour denotes the observed value (0.68) for source RM 102.}
\end{figure*}
\subsection{Red-to-blue optical Fe II blends ratio}
This part of Fe II emission is important for accurate modeling of the H$\beta$ zone, as recently stressed by \citet{popovic2023}. Proper modeling of the Fe II pseudo-continuum may strongly affect the conclusions about the dynamics of the BLR (like the presence or absence of the very broad line component to H$\beta$, and the role of the intermediate line region), and the modeling of the quasar main sequence \citep{panda2018, panda2019_WC, panda2019, Marziani_etal_2021}. 

We show contour plots for the ratio of red (5150-5350 \AA) to blue (4434-4684 \AA) parts of the optical Fe II in Figure~\ref{fig:red_blue_ratio_hbeta}. 
This ratio shows again the non-monotonic dependence of the ionizing photon flux but interestingly, almost no dependence on the density. 
The dependence on the microturbulence velocity is marginally noticeable.  
The measured value of this ratio in our test object is 0.68 (shown by red dashed contour) which could only be found for density $< 10^{10}$ cm$^{-3}$, ionizing flux $\sim$$10^{20}$ cm$^{-2}$ s$^{-1}$ and high microturbulence velocity. 

\begin{figure*}
\centering
\includegraphics[width=14cm, height=12cm]{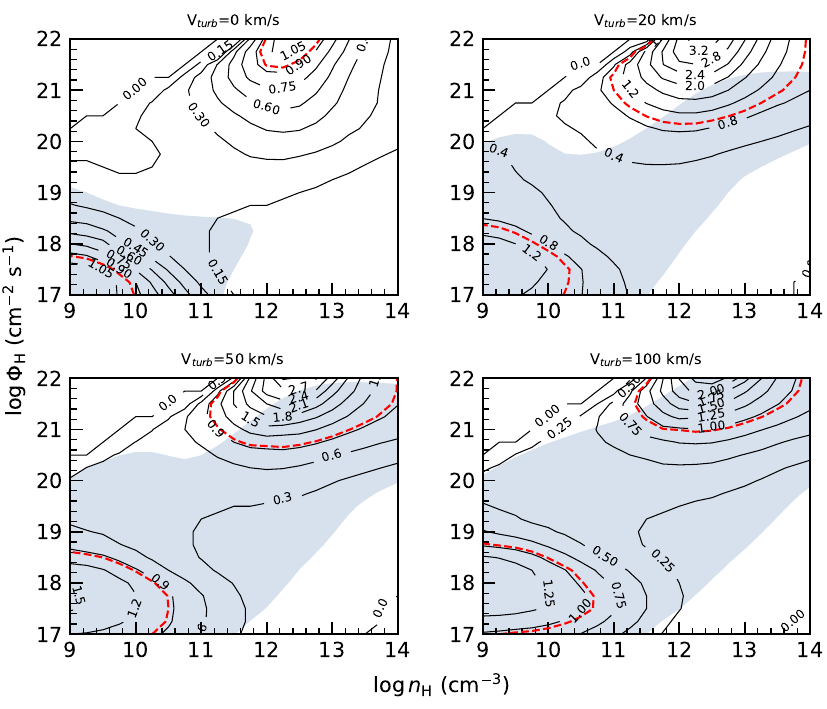}
\caption{\label{fig:fe2_hbeta_ratio}Contour plots for  R$_{\rm FeII}$ i.e. the ratio of Fe II 4434-4684 \AA \ to H$\beta$ ($\lambda$4861.32) fluxes for different microturbulence velocities. The shaded region represents the model parameter space that predicts the strength of optical blue Fe II emission consistent with the observed value, for a covering factor of 100\%. The red dashed contour denotes the observed value (0.95) for source RM 102.}
\end{figure*}

\begin{figure}
    \centering
\includegraphics[width=8cm, height=6cm]{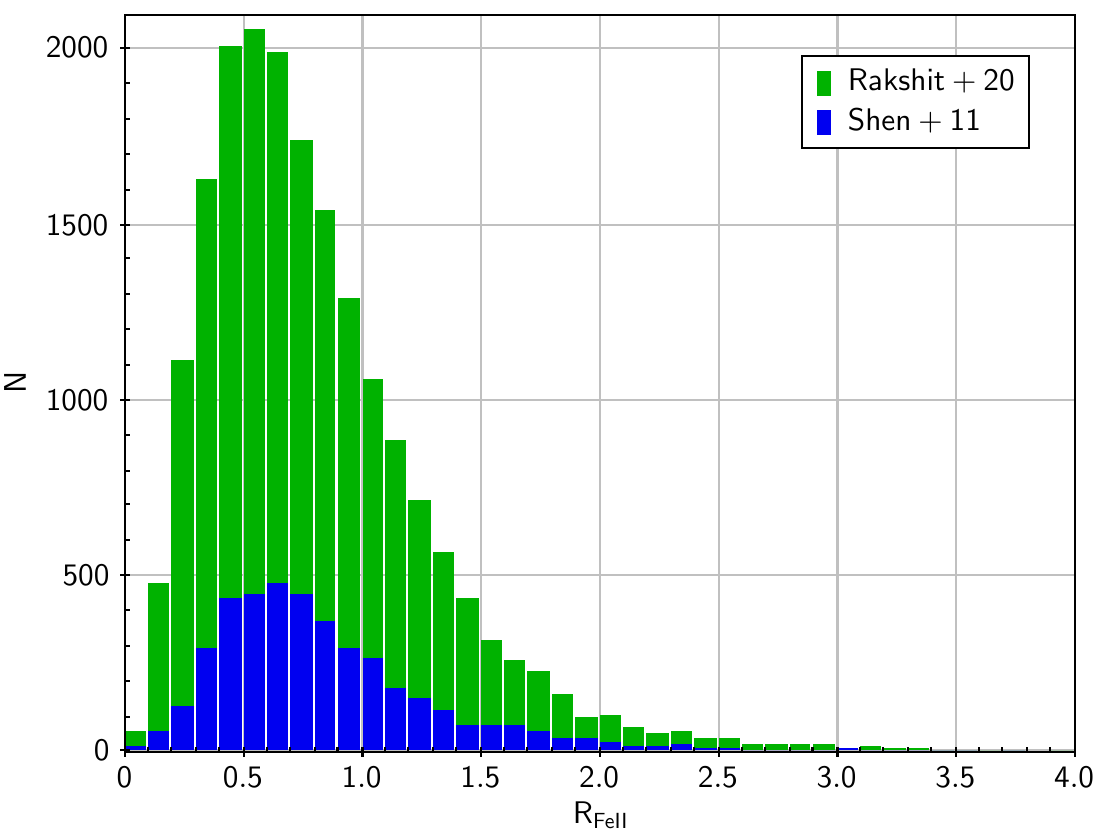}
    \caption{Histogram of the $R_{\rm FeII}$ distribution in SDSS quasar spectra based on \citet{shen2011} (blue) and \citet{rakshit2020} (green). }
    \label{fig:SDSS}
\end{figure}

\subsection{$R_{\rm FeII}$ in the optical band}\label{subsec:rfe2}
When computing the Fe II emission with \cloudy \ we also have, as a by-product, the intensities of other emission lines. The ratio $R_{\rm FeII}$ of the optical Fe II emission to H$\beta$ plays an exceptional role in the EV1 of the quasar sequence \citep{boroson1992, sulentic2000, shen2014, marziani2018, panda2018, panda2019, Du_Wang_2019, Panda_Marziani_2023, 2024ApJS..272...11P, 2024ApJS..272...13P}. We estimated $R_{\rm FeII}$ as the ratio of the integrated optical Fe II flux in the standard wavelength range 4434-4684 \AA \ to the flux of H$\beta$ at 4861.32 \AA. We present the corresponding contour maps in Figure~\ref{fig:fe2_hbeta_ratio}. 

These plots show that most of the map area corresponds to the value of $R_{\rm FeII}$ smaller than 1. Higher values, above 3, are only predicted for the intermediate values of the microturbulence velocity (20 - 30 km s$^{-1})$, and only for very high values of the ionizing photon flux ($\log \Phi_{\rm H}\,(\rm cm^{-2} s^{-1}) > 21$) which supports the earlier findings by \citet{panda2018} and \citet{panda2021}. 
Smaller values of $R_{Fe}$  correspond to the early A and B type quasars. The basic classification of quasars into class A and B, introduced by \citet{sulentic2000} was analogous to the older classification of Seyfert galaxies into Narrow Line Seyfert 1 (NLS1) class and other Seyfert 1 galaxies \citep{osterbrock1985}, only the limiting velocity was different (4000 km s$^{-1}$ vs. 2000 km s$^{-1}$), which reflected the systematic difference in the mass of the central black hole. 
Specific ranges of $R_{Fe}$ were used to explore full 2-D classification \citep[see e.g.,][]{marziani2018}. A-type quasars with $R_{\rm FeII}	\gtrsim 1$ were named Extreme Population A (xA) sources, and they were found to be applicable as standard candles for cosmology \citep{marziani2014,dultzin2020,marziani2023, Panda_Marziani_2023}. The xA sources have values of $R_{\rm FeII}$ between 1 and 2 \citep{marziani2018} but \citet{sniegowska2018} found one source with a large ratio. A few more such sources were reported by \citet{shen2011} but as discussed in \citet{sniegowska2018}, those measurements were not of high quality. 
Moreover, the sources with $R_{FeII} \gtrsim$ 2 are reported in the newer catalogs \citep[e.g.][]{negrete2018, rakshit2020, Rakshit_etal_2021, Wu_Shen_2022, 2024ApJS..272...11P}.

Assuming a simple power-law shape for the continuum, we observed R$_{\rm FeII}$ = 1.24 $\pm$ 0.7 for RM 102. When we modeled the continuum as a combination of a disk and stellar contributions, the revised value is R$_{\rm FeII}$ = 0.95 $\pm$ 0.35. This decrease is attributed to the inclusion of the host galaxy, which can lead to potential misclassification of a source as xA, as demonstrated by \cite{bon2020}. However, both measurements are consistent within uncertainties and meet the criterion for xA sources (R$_{\rm FeII} \geq 1 $). Additionally, \cite{shen2019} reported an Eddington ratio of $\lambda_{\rm Edd}$ = 0.329  for this source. Both the Eddington ratio and R$_{\rm FeII}$ measurements confirm that RM 102 belongs to the xA class of AGN. 

We noticed that the observed R$_{\rm Fe II}$ value for RM 102 falls within two distinct regions of the contour plots in Figure \ref{fig:fe2_hbeta_ratio}: one corresponding to low-density and low-ionizing flux and the other to high-density and high-ionizing flux. Previous studies on UV emission line ratios in xA sources \citep[e.g.][]{Negrete2012,panda2019} have favoured the possibility of a high-density ($\log n_{\rm H}$ (cm$^{-3}$) = 12 - 13) BLR for these objects.
We, thus, may conclude, from Figure \ref{fig:fe2_hbeta_ratio}, that xA sources require a high-density BLR ($\log n_{\rm H}$ (cm$^{-3}$) = 12 - 13) and very high ($\log \Phi_{\rm H}\,(\rm cm^{-2} s^{-1}) > 20$) ionizing flux. However, our templates were composed of solar metallicity. Some authors argued \citep[e.g.][]{mlma18,panda2019,sniegowska2021, panda2021, Marziani_etal_2024, 2024arXiv240504456F} that these sources, in addition, require highly super-solar metallicity. 

We compared these predictions with the statistical distribution of the $R_{\rm FeII}$ reported based on SDSS quasar studies. We used two data sets: older massive fits to the SDSS quasars, based on data release 7 (DR 7), were reported by \citet{shen2011}. Newer results, based on DR 14, were reported by \citet{rakshit2020}. The results are based on the spectral decomposition of the Fe II emission integrated within the traditional range 4434-4684 \AA\ \citep{boroson1992}. In both cases, we selected only the sources with reliable spectra decomposition by requiring that the relative error of the FWHM of H$\beta$ and the $R_{\rm FeII}$ measurement are below 20\%. The resulting histograms, shown in Figure~\ref{fig:SDSS}, consistently show the peak of the $R_{\rm FeII}$ values at about 0.6 in both samples. There are very few objects with values above 4.0 which we ignored for better clarity of the plot. Such values are present either in the region of higher density and high ionizing photon flux, above $\sim 10^{20}$ cm$^{-2}$ s$^{-1}$, or low ionizing photon flux and low density (lower left corner of plots in the Figure~\ref{fig:fe2_hbeta_ratio}). The first location seems more likely as it is smoothly connecting to even larger values of $R_{\rm FeII}$ than the average value, and the observational histogram also smoothly extends to higher values. However, we should stress that the observational results are not based on the Fe II emission models presented in this paper. Results of both \citet{shen2011} and \citet{rakshit2020} are based on the use of the optical Fe II template from \citet{boroson1992}.

\begin{table*}
    \centering 
    \caption{\label{tab:metallicity_effects} The model predicted ratios of Fe II emission for various abundances,  with a fixed microturbulence velocity of 100 km/s, in different wavebands considered in this work.}    
    \begin{tabular}{cccccc} \hline\hline
Abundance (in  $Z_{\odot}$) & $\log \Phi_{\rm H}$,$\log n_{\rm H}$ & UV Fe II/optical Fe II & UV red/blue wings  & Optical red/blue blends  & R$_{\rm FeII}$ \\ \hline
1   & 18,10   & 4.96  & 1.59 & 1.80 & 1.20      \\
    & 20.5,12 & 11.82  & 1.23 & 1.03 & 0.60    \\
5   & 18,10   & 4.13   & 1.63  & 1.91 & 1.59     \\
    & 20.5,12 & 6.62   & 1.49  & 0.90 & 1.51     \\
10  & 18,10   & 4.17  & 1.64   &  1.91   &  1.47  \\
    & 20.5,12 &  5.33   & 1.56   & 0.86  &  2.04   \\  
\hline
    \end{tabular}
\end{table*}

\begin{figure*}
\centering
\includegraphics[width=14cm, height=12cm]{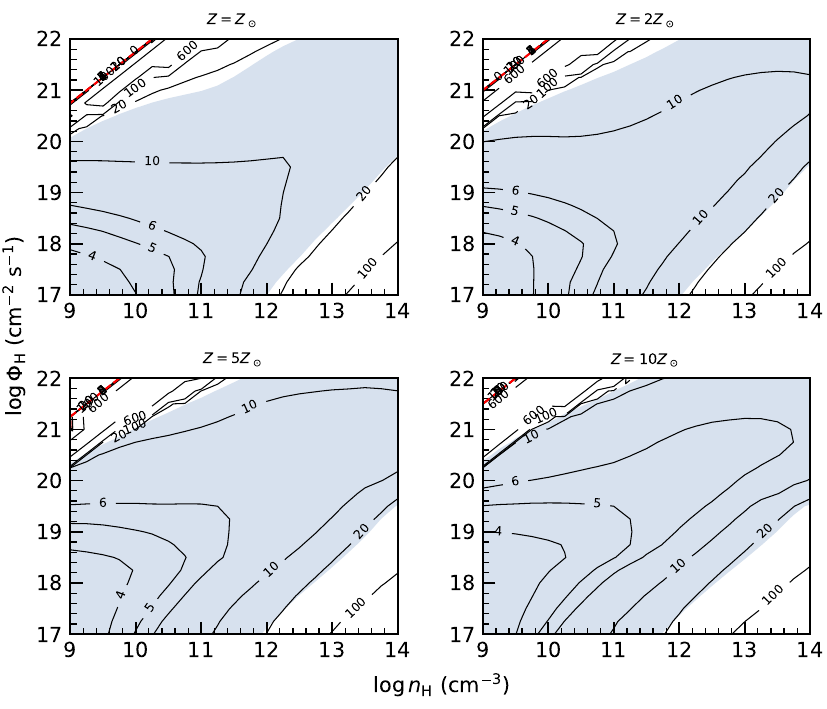}
\caption{\label{fig:vary_Z_uv_optical_fe2_ratio} Contour plots for the ratio of UV Fe II to optical Fe II emission for microturbulence velocity V$_{turb}=100$ km/s and different metallicities. The shaded region represents the parameter space reproducing the optical broadband Fe II emission strength, assuming a covering factor of 100\%, in test object RM 102. 
}
\end{figure*}

\begin{figure*}
\centering
\includegraphics[width=14cm, height=12cm]{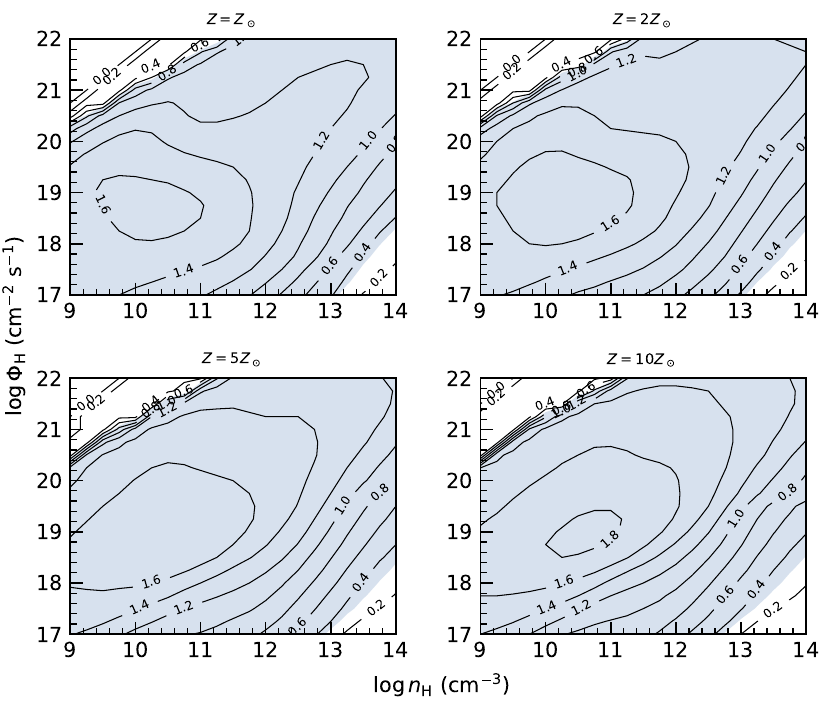}
\caption{\label{fig:vary_Z_red_blue_ratio}Contour plots for the ratio of red (2800-3090 \AA) to blue (2650-2800 \AA) part of the UV Fe II spectra for microturbulence velocity V$_{turb}=100$ km/s and different metallicities. The shaded region denotes the model parameter space that predicts the EW of UV red Fe II wing consistent with the observed value for a 100\% covering factor.
}
\end{figure*}

\begin{figure*}
\centering
\includegraphics[width=14cm, height=12cm]{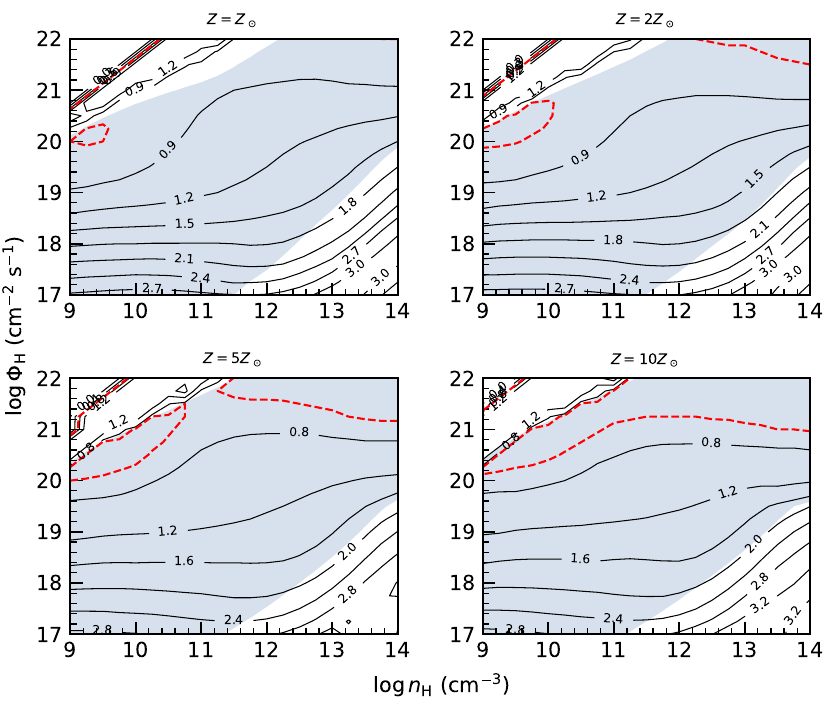}
\caption{\label{fig:vary_Z_red_blue_ratio_hbeta}Contour plots for the ratio of red (5150-5350 \AA) to blue (4434-4684 \AA) part of the optical Fe II spectra for microturbulence velocity V$_{turb}=100$ km/s and different metallicities. The shaded region denotes the model parameter space that predicts the strength of optical blue Fe II emission consistent with the observed value, for a covering factor of 100\%.
The red dashed contour denotes the observed value (0.68) for source RM 102.}
\end{figure*}

\begin{figure*}
\centering
\includegraphics[width=14cm, height=12cm]{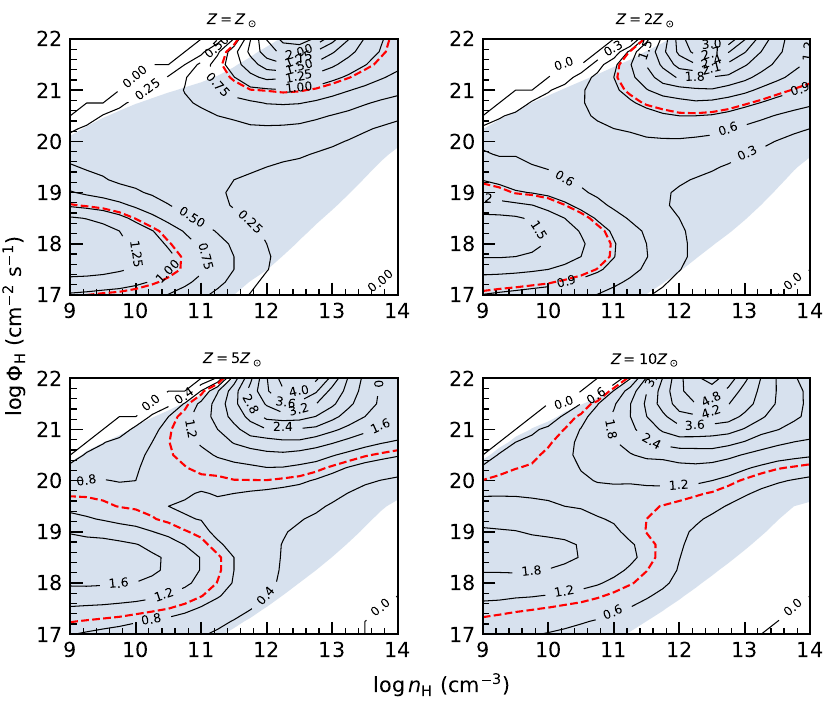}
\caption{\label{fig:vary_Z_fe2_hbeta_ratio}Contour plots for  R$_{\rm FeII}$ i.e. the ratio of Fe II 4434-4684 \AA \ to H$\beta$ ($\lambda$4861.32) fluxes for microturbulence velocity V$_{turb}=100$ km/s and different metallicities. The shaded region represents the model parameter space reproducing the strength of optical blue Fe II emission, assuming a covering factor of 100\%, consistent with the observed value for the test object RM 102. 
The red dashed contour denotes the observed value (0.95) for source RM 102.}
\end{figure*}

\subsection{The effect of metallicity}
To investigate the effect of metallicity on Fe II emission, we rebuild our models for abundances equal to  2, 5, and 10 times solar abundances, with a fixed microturbulence velocity of 100 km/s. We plot contours for the ratios estimated in the previous sections in Figures \ref{fig:vary_Z_uv_optical_fe2_ratio}-\ref{fig:vary_Z_fe2_hbeta_ratio} for various abundances. The values of these ratios at two arbitrary positions are given in Table \ref{tab:metallicity_effects} to quantify their variations with abundances. 

The UV Fe II to optical Fe II contours, shown in Figure \ref{fig:vary_Z_uv_optical_fe2_ratio}, exhibit marginal changes with abundances. The ratio decreases with increasing abundance at ($\log \Phi_{\rm H}, \log n_{\rm H})$=(20.5,12), however, no such trend is seen at lower incident flux, and lower density ($\log \Phi_{\rm H}, \log n_{\rm H})$=(18,10). \cite{verner2003} investigated the effect of varying abundance on the ratio of UV Fe II to optical Fe II and found that the ratio decreases from 8 to 5.6 when abundance increases from solar to 10 times solar abundance for $\log \Phi_{\rm H}\,(\rm cm^{-2} s^{-1}) = 20.5, \log n_{\rm H}\, (\rm cm^{-3}) = 9.5$ and a microturbulence velocity of 1 km/s. They concluded that the optical Fe II emission increases more than the UV Fe II emission when the abundance increases which is supported by the findings from \citet{panda2019_WC}.

Figure \ref{fig:vary_Z_red_blue_ratio} shows the contours for the ratio of UV wings, whereas Figure \ref{fig:vary_Z_red_blue_ratio_hbeta} shows those of optical blends. As seen from the Figures and Table \ref{tab:metallicity_effects}, both ratios are essentially unaffected by the abundance. 

The contours for R$_{\rm FeII}$ are plotted in Figure \ref{fig:vary_Z_fe2_hbeta_ratio} which shows small variations with abundance. 
The ratio increases with increasing abundance at high ionizing photon flux and high density ($\log \Phi_{\rm H}, \log n_{\rm H})$=(20.5,12), but no such trend is seen at ($\log \Phi_{\rm H}, \log n_{\rm H})$=(18,10).

\subsection{Fe II templates}
\begin{figure}
    \centering
    \includegraphics[width=8cm, height=6cm]{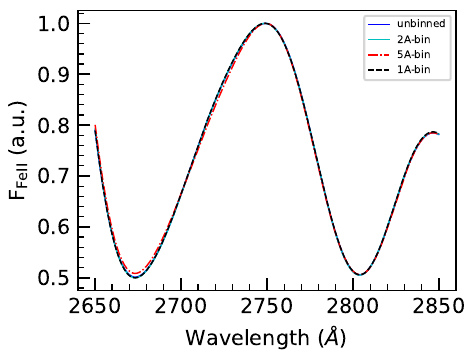}
    \caption{Plot showing  Fe II emission (in arbitrary units) over a wavelength range of 2650-2850 \AA \ around the UV Mg II line to justify the choice of 2 \AA \ bin size. The larger bin size (5 \AA) already shows a visible departure in the smoothed shape from the denser grid.}
    \label{fig:bin_test}
\end{figure}
For convenient use in future spectral fitting, we created a new family of Fe II templates. They come from the simulations as described in Section \ref{sec:method}, i.e. those are 2646 models covering the range of the local density, ionizing photon flux and microturbulence. Output files from \cloudy \ are usually large, and data fitting does not require full resolution available in these output files since most of AGN have broad iron components. Therefore, we binned the output spectra using the 2 \AA~bins. We tested whether this does not lead to a considerable problem in data fitting by assuming that the AGN spectrum comes from the medium with the velocity dispersion 2000 km s$^{-1}$ and we compared the Fe II spectrum in the 2650 - 2850 \AA~ range without any prior binning, and with a bin size of 1, 2 and 5 \AA. The result is shown in Figure~\ref{fig:bin_test}. Up to the bin size of 2 \AA\ there is no net difference in the broadened spectra but the bin size of 5 \AA\ seems to imply already a systematic difference, so we adopted the 2 \AA\ bin size optimizing between the accuracy and the size of the file. 

Additionally, we include templates for the inward and outward Fe II emission, which are split into 1000 logarithmic wavelength bins, each $\sim$ 584 km s$^{-1}$ wide, between 1000 and {7000 \AA}. The complete set of Fe II templates for solar metallicity is available through zenodo\footnote{\url{https://zenodo.org/doi/10.5281/zenodo.10532690}} and GitHub\footnote{\url{https://github.com/Ashwani-88/Fe2_template}} links. Each template corresponds to a single set of cloud parameters such as local density (n$_{\rm H}$), ionizing photon flux ($\Phi_{\rm H}$), and microturbulence ($V_{\rm turb}$). In the next Section, we compare our several Fe II templates based on single cloud parameters with observed Fe II spectra to comprehend the physical properties of the BLR cloud emitting Fe II emission. However, it is possible that a combination of clouds having different physical conditions can be responsible for the observed Fe II spectra as well as the other strong emission lines. Our Fe II templates can be combined using any multi-cloud model such as the Locally Optimized Cloud (LOC) model suggested by \cite{baldwin1995} to compare the observed Fe II spectra of quasars.

Reverberation mapping studies indicate that the broad-line region (BLR) is radially extended and likely exhibits vertical stratification, along with a range of local densities due to thermal instabilities and the cloud formation process. Our publicly available set of Fe II templates is thus envisioned as a good starting point for further modelling, using LOC or other BLR models which include stratification. However, such modelling is beyond the scope of our current objective.

\subsection{Testing the templates against the complete spectra of the two test objects}

We test how the new Fe II templates compare with the observed AGN spectra instead of using just integrated spectral bands. 
In the case of HE 0413-4031, only narrow UV band data are available while the spectrum of SDSS~RM102 continuously covers the broad UV-to-optical band.

\subsubsection{HE 0413-4031}
\label{sect:HE0413}
The observed spectrum of HE 0413-4031 covers a narrow range around the Mg II line but the spectral shape did not show any strong variability apart from the normalization of the spectrum. We thus constructed the mean spectrum of this quasar by combining all 32 available data sets for a better S/N ratio. The shape of the Mg II line in this source is particularly simple, well fitted as a single kinematic component of the Lorentzian shape. 

We fit all the parameters together (slope and normalization of the power law, normalization and the broadening of the Fe II template, the position, normalization, and FWHM of the Mg II) since in case of such a short data range we do not have any possibility to find parts of the spectrum unaffected by Fe II and fit first the underlying power law. 

\begin{figure}
    \centering
    \includegraphics[width=8cm, height=6cm]{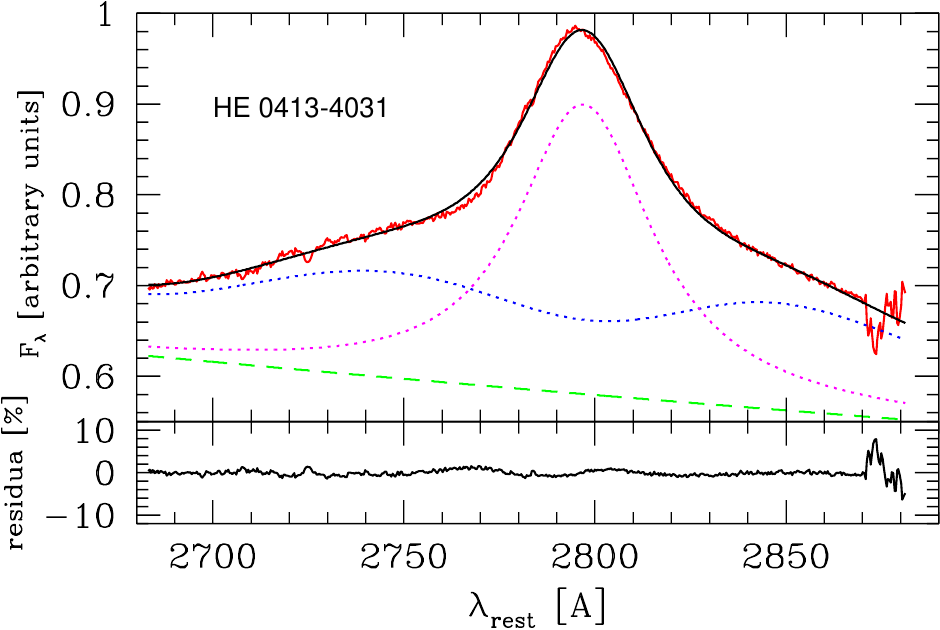} \\
    \includegraphics[width=8cm, height=6cm]{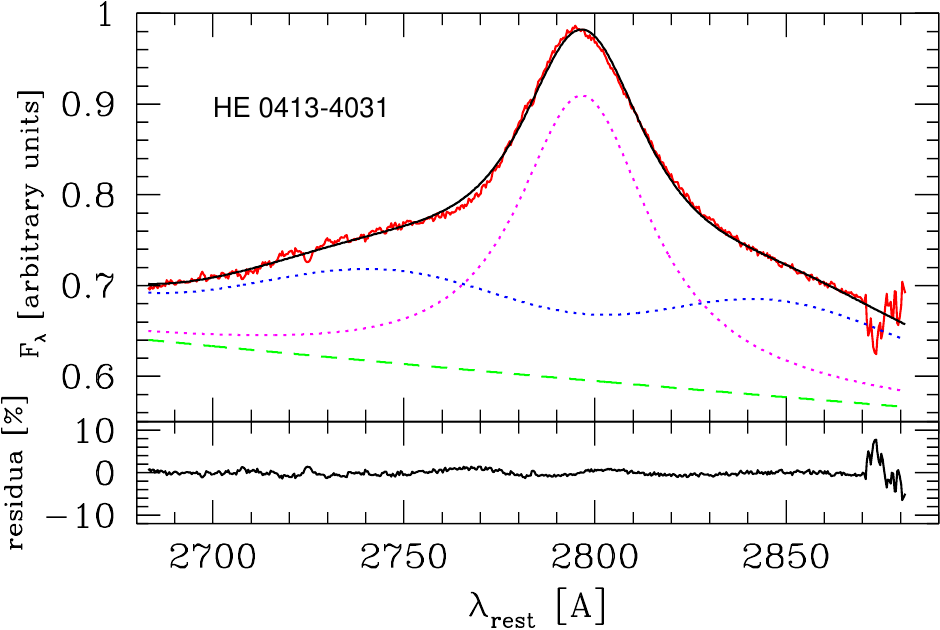}\\
    \includegraphics[width=8cm, height=6cm]{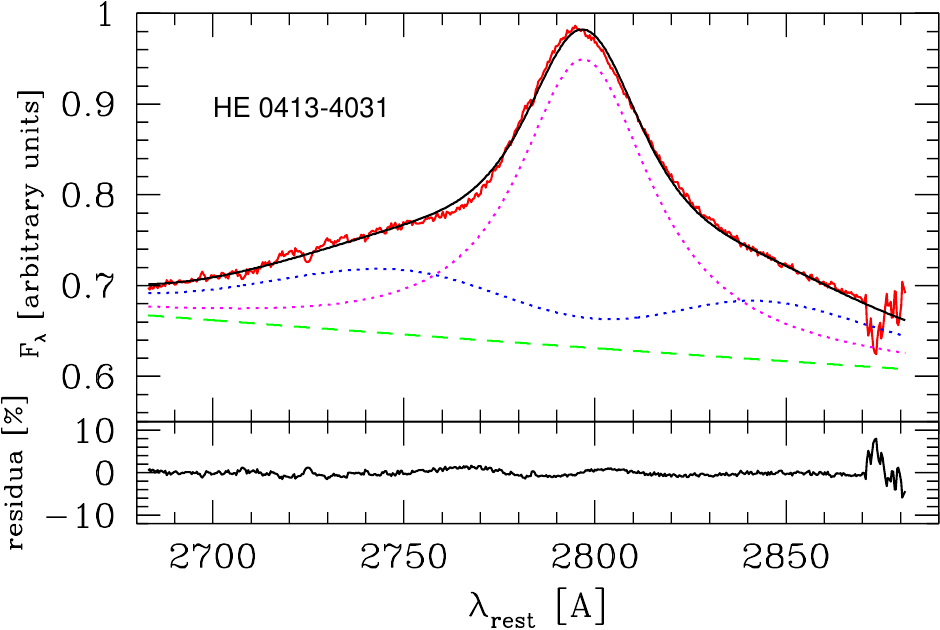}
    \caption{\label{fig:HE_0413}The exemplary fits to the SALT spectrum with the Fe II template phi19.0-nH12.0-m50.dat, FWHM = 5400 km s$^{-1}$ (upper panel), phi18.25-nH12.0-m50.dat, FWHM = 5400 km s$^{-1}$ (middle panel), and phi18.0-nH12.0-m50.dat, FWHM = 5800 km s$^{-1}$ (lower panel). The middle panel shows the best fit.}
    
\end{figure}

The narrow spectral range ($\lambda_{\rm rest}\sim$ 2700 - 2900 \AA), centered around the broad Mg II line, makes the fits not strongly constraining the Fe II templates but the fits are very satisfactory.  The results of the spectral fitting are shown in Figure \ref{fig:HE_0413}. The best solution corresponds to $\log \Phi_{\rm H}\,(\rm cm^{-2} s^{-1}) = 18.25$, shown in the middle panel of Figure \ref{fig:HE_0413}. The best fit FWHM of the Fe II (5400 km s$^{-1}$) is broader than the FWHM of Mg II (4280 km s$^{-1}$) but it comes with a large error. Higher microturbulence velocity (100 km s$^{-1}$) gives fits with somewhat higher reduced $\chi^2$, and pushes the FWHM of Fe II toward even higher values, at the same time increasing the Fe II intensity. This results from considerable degeneracy between the Fe II normalization and the underlying power law. A lower value of the microturbulence velocity (20 km s$^{-1}$) gives narrower Fe II lines but the quality of the fit is much worse.

For comparison, we show in Figure~\ref{fig:HE_0413} two other fits, for higher ($\log \Phi_{\rm H}$ (cm$^{-2}$ s$^{-1}$) = 19) and lower ($\log \Phi_{\rm H}$ (cm$^{-2}$ s$^{-1}$) = 18) ionizing photon fluxes in the upper and lower panels of Figure \ref{fig:HE_0413}, respectively. The fit quality is good in all these cases, the value of $\chi^2$ measuring the fit quality was higher only by $\sim 1.0$ for the upper and the lower panel than for the middle panel), and the value of the EW of the Mg II line is similar (31.2 \AA, 30.2 \AA~ and 28.4 \AA), but slowly rising towards larger ionizing photon fluxes. We also calculated EW(Fe II) but only in the fitted range. EW(Fe II) is more affected by the change of the ionizing flux (the corresponding values are 37.9 \AA, 31.2 \AA~, and 15.9 \AA), and all these values are larger than achieved with the use of different templates by \citet{zajacek2023}. Such a large change is related to a corresponding change in the slope and normalization of the underlying power law. 

The chosen fit density ($\log n_{\rm H}$ (cm$^{-3}$) = 12) is consistent with the best fit theoretical template d12-m20-20-5.dat selected from the collection of \citet{bruhweiler2008} in our previous papers, including  \citet{prince2023}. However, the ionizing photon flux is now lower by more than two orders of magnitude than in the template; $\log \Phi_{\rm H}$ (cm$^{-2}$ s$^{-1}$) = 18 in the model versus $\log \Phi_{\rm H}$ (cm$^{-2}$ s$^{-1}$) = 20.5 in the template.

We compared the requested value for the ionizing flux from Fe II modeling with the expected ionizing flux based on the SED of this object and the Mg II time delay of 302.6 days for Mg II \citep{zajacek2020}. The obtained ionizing photon flux from the SED best fit presented in Figure~9 of \citet{zajacek2020} gave  $\log \Phi_{\rm H}\,(\rm cm^{-2} s^{-1}) = 20.05$, although the spectrum appears to decrease rapidly in the far UV band. This ionizing flux is higher by two orders of magnitude than implied by the current fits, and consistent with the old \citet{bruhweiler2008} template used in \citet{zajacek2020} and \citet{prince2023}.
The narrow available spectral range on one hand guarantees that a range of templates fits well the spectrum but it opens a question of how reliable is such a decomposition. 

\begin{figure}
    \centering
    \includegraphics[width=18cm, height=16cm]{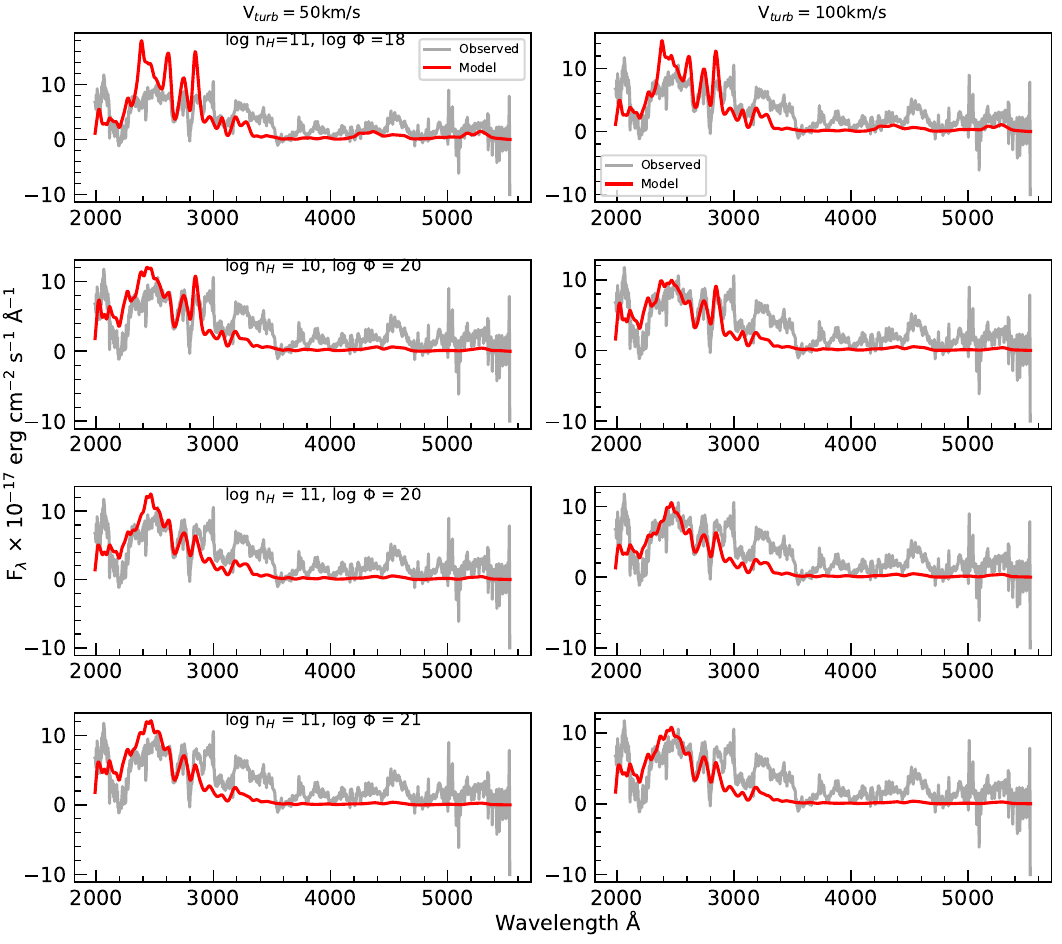}
    \caption{\label{fig:sample_spectra}A comparison of observed and model-predicted Fe II spectra over a wavelength range 1989-5530 \AA \ for different models. Plots in the left column are for microturbulence 50 km/s, while those in the right column are for the same models but for microturbulence 100 km/s.}
    \end{figure}
\subsubsection{RM 102}
\label{sect:RM102}

We estimated the ionizing photon flux from this source at the distance corresponding to the BLR position as given by the H$\beta$ line delay. However, the SDSS spectrum does not cover the spectral range above 1 Rydberg, needed for this purpose. We thus extrapolated the spectrum of RM102 using the AGN SED template from \cloudy. The observed spectrum is only marginally bluer in the optical band. Matching the two spectra at the UV part we obtained the ionizing photon flux of  $2.67 \times 10^{20}$ cm$^{-2}$ s$^{-1}$. We also tested another way of extension, using the HST composite spectrum of \citet{zheng1997} which covers the rest frame range from 350 \AA~ up to 2200 \AA, and they give the broken power law fit. Matching this broken power law to the shortest wavelengths with our spectrum we obtained the ionizing photon flux of $1.35 \times 10^{20}$ cm$^{-2}$ s$^{-1}$. The two values are comparable.
However, some composites predict a stronger decrease of the quasar far-UV flux towards shorter wavelengths \citep[see ][and the references therein]{cai2023}.  

\begin{figure*}
    \centering
\includegraphics[width=18cm, height=10cm]{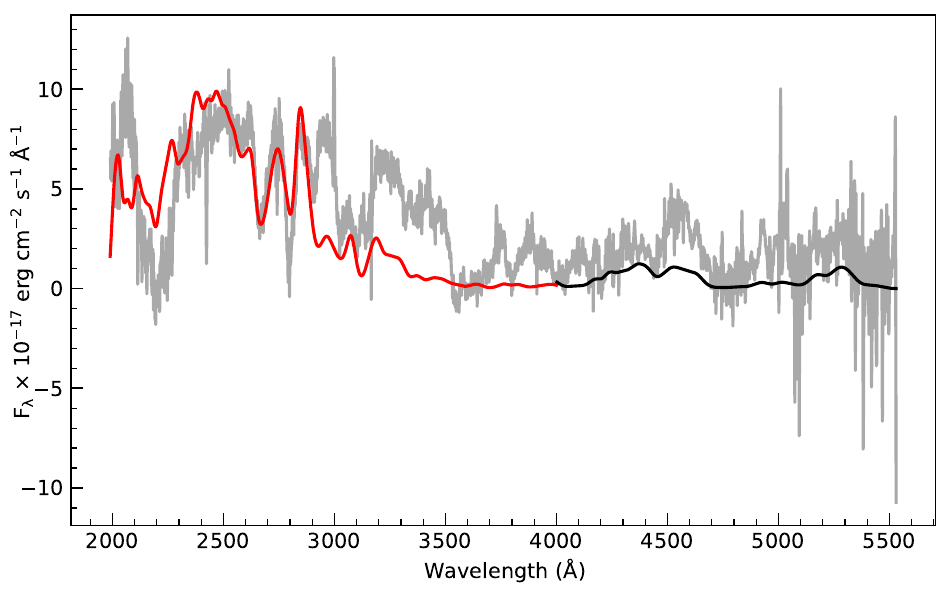}    
    \caption{A comparison of the observed Fe II emission in RM~102 (grey curve) with separate theoretical templates for UV and optical parts (see text). The red curve denotes the Fe II flux in the wavelength range 1989-4000 \AA \ for $\log \Phi_{\rm H}\,(\rm cm^{-2} s^{-1}) = 20$ and a covering factor of 30$\%$, while the black curve represents the Fe II emission in the wavelength range 4000-5530 \AA \ for $\log \Phi_{\rm H}\,(\rm cm^{-2} s^{-1}) = 19$ and a covering factor of 50$\%$. The gas density is fixed at $10^{10}$ cm$^{-3}$ and the microturbulence velocity is 100 km s$^{-1}$.}
    \label{fig:RM102_templates}
\end{figure*}

To compare our model-predicted Fe II spectra with the observed spectra of RM 102, we plotted sample spectra from different models and for microturbulence velocities 50 and 100 km/s in Figure \ref{fig:sample_spectra}. These spectra include a covering factor of 30\% and are convolved with a Gaussian function having FWHM = 4000 km s$^{-1}$. 
Figure~\ref{fig:sample_spectra} implies that a single Fe II template is unlikely to fit well the whole spectrum. In the UV part, the model-predicted spectra provide a somewhat better match to the observed spectra but in the optical region, the model-predicted Fe II emission is much weaker compared to the observed Fe II.
The metallicity is not affecting the ratios strongly enough. We thus set the density at the likely value of $10^{10}$ cm$^{-3}$, and we took a different approach to fit the observed spectra. We allowed for a discontinuity in the ionizing flux at 4000 \AA, where the Fe II emission is very low. We used a high ionizing flux ($\log \Phi_{\rm H}\,(\rm cm^{-2} s^{-1}) = 20$) in the UV band and a low ionizing flux ($\log \Phi_{\rm H}\,(\rm cm^{-2} s^{-1}) = 19$) in the optical band. Separate fitting of the UV and optical Fe II is usually done when modeling a specific source.  We also chose different covering factors in UV (30\%) and optical (50\%) regions. The need for a higher covering factor in the optical part has been suggested in several studies \citep{2009NewAR..53..140G,ferland2009}.
The result of this approach is shown in Figure~\ref{fig:RM102_templates}.
None of these attempts are fully satisfactory. However, the discontinuous model seems to work much better. Still, in both cases, we face a considerable mismatch between the model and the data in the region 3000 - 3500 \AA. Semi-empirical models do include considerable emission there (see bottom panel of Figure~\ref{fig:bcz}, dotted line). The spike close to $\sim$ 2000 \AA \ in the data is most likely a calibration issue. 

\section{Discussion}
\label{sect:discussion}
We used the most recent version of the code \cloudy \ C23.00 to construct the grid of theoretical Fe II templates covering the broad UV to NIR range from 1000 \AA \ to 10000 \AA. The templates represent single-zone emitters, parametrized by the local density and ionizing photon flux. We consider a range of microturbulence velocities and for a fixed microturbulence velocity of 100 km s$^{-1}$ we calculated the results for different metallicities. For the other microturbulence velocity the considered metallicity is solar. 

The advantage of the theoretical templates is that they predict the Fe II emission underlying the strong emission lines while purely observational templates, modeling first the lines, force low Fe II emission at the position of lines like H$\beta$ or Mg II. In addition, the construction of the observational templates is thus most successful for the objects with extremely narrow emission lines like I Zw 1 \citep{vestergaard2001} or Mrk 493 \citep{park2022}. Line widths of these spectra imply a high Eddington ratio, strong Fe II emission, and the incident SED shape not necessarily representative of the majority of quasars. 

The availability of the Fe II templates in the full spectral range will address better the issue of the production of the Fe II in the BLR. In the case of the observational data covering a narrow spectral range, the templates allow us to fit the data and obtain an estimate of the density and the ionizing photon flux in the BLR, as we showed in Section~\ref{sect:HE0413}. However, our attempt to use these templates for a quasar spectrum covering both optical and UV ranges was unsuccessful (see Section~\ref{sect:RM102}). The problem seems generic since the selected quasar is rather a typical object, and the issue of the UV to optical ratio in the models and the data has been present in the literature for many years.

Our template fluxes are properly normalized so we can check the equivalent width of the Fe II emission in various bands corresponding to 100 \% covering factor of the central source by the BLR region. Overall, the Fe II emission strength of the irradiated faces of the clouds is large enough to represent the actual emission when the realistic covering factor is included. However, in most bands, it requires relatively high microturbulent velocity, close to 100 km s$^{-1}$. Such high microturbulence velocity was discussed in some previous works \citep[e.g.][]{baldwin2004, panda2018, panda2019_WC, sarkar2021}. A microturbulence of this magnitude must reflect the dispersion of kinematic velocity rather than the thermal motion of the gas. 
This is not unexpected in the case of a turbulent medium. The medium, most likely formed as a clumpy wind from the disk \citep[e.g.][]{murray1995,czhr2011,elvis2017} shows systematic Keplerian motion of a few thousand $\sim$ km s$^{-1}$ but clump formation or failed wind approach can introduce the velocity field being a fraction of this Keplerian speed, and affect the transparency of the medium. Thus the medium represented by a combination of directly irradiated and not irradiated clouds in roughly equal proportion can well represent the BLR. Additionally, the magnetic field likely present in the BLR can lead to the development of internal motions at a level of even a few hundred km s$^{-1}$ \citep{2012MNRAS.425.3172K}.

 We also studied separately the dark (not directly irradiated) cloud sides since some inflow models and special geometries may suggest that we mostly see the dark sides of the clouds \citep[e.g.][]{ferland2009,wang_inflow2017}. If clouds are few and randomly distributed around the central source, then we will see clouds at all angles to their orientation toward the source. The entire contribution summed up is well represented by the isotropic emission. If cloud distribution is flattened, but our line of sight is not passing through the cloud region the shielded sides may contribute a little stronger to the total emission in terms of the effective surface but most of the cloud emission comes from the bright sides. Such expectations were given, for example, by \citet{czhr2011} in the context of the time delay. However, the bright sides contribute more to the isotropic part, therefore overexposure of the cloud's bright sides may not have a significant impact on Fe II emission. The dominance of the shielded material changes the picture considerably. 
\citet{ferland2009} connected this effect to the Fe II emission in the optical band showing predominantly the redshift \citep{Hu_Chen2008} and thus implying the inflow. Also in \citet{panda2018}, in Section 4.3, they discussed the issue that the optical Fe II is produced predominantly in the unilluminated side of a cloud, and the dominance of the unilluminated side helped to reproduce $R_{\rm Fe II}$ ratio without metallicity enhancement. The requested geometrical setup is not simple since the line of sight does not cross the BLR. It may require a highly ionized absorber, between the black hole and the BLR, almost `invisibly' shielding the well-irradiated bright sides of clouds. However, we show now that if dark sides statistically dominate, the strength of Fe II emission is too low to reproduce the overall Fe II emission in typical quasars. 

Moreover, when confronting the templates with the broadband optical/UV data with the test object RM 102 we could not well represent the ratios of the UV-to-optical band, despite the use of the newest atomic data. The problem was already discussed by \citet{Joly1981J,verner2003,baldwin2004}. \citet{baldwin1995} used the Locally Optimized Cloud (LOC) model which allowed for a combination of clouds with a range of densities and distances to solve the issue and concluded that UV and optical emission come from two separate regions of different properties. But 
the radially very extended emission is unlikely. \citet{KDP2015} argued that there is a kinematic correlation between the UV and the optical Fe II emissions, although they reported considerable positive shift of the UV Fe II for the systemic redshift for the majority of their sources (their Fig.~10). The radius-luminosity relation discussed by \citet{zajacek2023} for both spectral ranges looks roughly similar, so the two regions cannot be considerably shifted. The problem of matching the theoretical templates over the full optical to UV spectral range is not likely due to the use of a single-zone approximation, particularly as some of the papers mentioned above used a multi-zone approach. 

However, we should note that there may also be a problem on the observational side. Despite the selection of the object with a high S/N ratio, the determination of the strength of Fe II emission is not perfect, particularly in the optical band where we also have some starlight contribution. We do not think it is the main source of the problem, but certainly, future applications of our new templates using the LOC model and more objects can shed light on the issue.

There are thus two remaining issues that can solve the problem and could be done in principle but they were not modeled in this paper since our primary goal was to provide a set of theoretical templates for further use. We do not discuss below the issues which are known but are far beyond the modeling possibilities. These are the dynamical aspects of cloud evolution: formation and destruction, including the collisions with the accretion disk. For a short summary and the references, see Appendix in \citet{muller2022}.

The first weakness in the simulation is the use of a constant density assumption. Such an approach is widely employed, yet it is not physically warranted. The higher temperature of the illuminated face of the cloud and the considerably lower temperature of the shielded side would create an enormous pressure gradient which could disrupt the cloud on a dynamical timescale. Instead, the illuminated cloud must be in pressure equilibrium which also takes into account the radiation pressure \citep{rozanska2006,baskin2014}. Some models also suggest the requirement of magnetic pressure for the stability of BLR clouds \citep[e.g.][]{2012MNRAS.425.3172K}.
Overall, the bright face has a much lower density, and the density is rising towards the unilluminated side. Clouds in pressure equilibrium were considered in several papers. Some of these computations were done in the context of the warm absorber \citep{rozanska2006,adhikari2015,2019adhikari} but this medium may be the same as BLR.  The change of the density and temperature is not like a power law but rather like a rapid transition (see e.g. Fig.~1 of \citealt{baskin2014}, or Fig.~3 of \citealt{2019adhikari}), so the LOC model would not reproduce it well, and eventually, a discontinuous jump in density and ionizing photon flux might be a better approximation. Computations of clouds under constant pressure are possible within \cloudy \ environment, and this could be done in the future. Constant pressure assumption will allow for an approximate co-existence of material at strongly different local densities which would be forced by the pressure balance and the temperature gradient, although this would never represent yet the more complicated mixture of conditions which could form when the dynamical aspect of cloud formation/destruction is taken into account. 

The second problem is the heating mechanism. Already the early papers asked the question of whether the Fe II emission forms as a result of the irradiation by the central source or due to mechanical heating. For example, \citet{Joly1981J} argued that either mechanical heating is the dominating channel of production, or the clouds must be very optically thick, in which case the irradiation flux is thermalized and the mechanism is similar to the mechanical heating. 

The possibility of measuring the Fe II delay in the optical and UV bands contradicts the major collisional origin of the emission, but some contributions cannot be ruled out. Mechanical heating was discussed by \citet{verner2003}. The UV/optical ratio obtained in their paper is similar to what we measured from RM102 in the case of collisional heating, but radiative heating gave much higher values, 8 - 11, depending on the number of transitions. These values were obtained under a microturbulence velocity of 1 km s$^{-1}$. They noted that Fe II optical emission flux is more sensitive to abundance than the Fe II (UV) band. Conversely, the Fe II (UV) band is more sensitive to microturbulence than the Fe II optical band. Also \citet{baldwin2004} considered an option of mechanical heating but they overall favoured the radiative heating model with strong microturbulence. Such a model, however, requires two separate emitting regions for optical and UV, as mentioned above. Therefore, the combination of mechanical and radiative heating might be an interesting motion for future studies.

\section{Summary}\label{sec:summary}
One of the most prominent characteristic features of the quasar spectrum is the iron emission. In this work, we examine the properties of Fe II emission using \cloudy \ version C23.00 and generate new theoretical Fe II templates that can be used to model Fe II emission in the quasar spectra covering the wavelength range 1000-10000 \AA. We study the impact of microturbulence and metallicity on the strength of Fe II emission and the shape of the iron emission and compare our model predictions with observational data. The primary outcomes of our study are outlined below.
\begin{enumerate}
    \item The unilluminated sides of the clouds do not produce enough flux to represent the observed Fe II emission
    \item A high microturbulence velocity is required to match the observed EW of Fe II and the shape of the UV Fe II emission.  
    \item The updated atomic database available in the latest version of \cloudy \ C23.00 could not resolve the long-standing problem of a mismatch between the expected and observed values of Fe II UV to optical ratio under the single zone approximation.
    \item Microturbulence has less impact on the UV to optical Fe II emission ratio.
    \item An additional apparent velocity shift of up to 1000 km/s can be added to the spectrum by microturbulence.
    
      
    \item The ratios of Fe II emissions are not significantly affected by the metallicity. 
\end{enumerate}

\begin{acknowledgments}
We are grateful for the referee's insightful remarks and suggestions.
The project was partially supported by the Polish Funding Agency National Science Centre, project 2017/26/A/ST9/00756 (MAESTRO 9). AP acknowledges funding from the Chinese Academy of Sciences President’s International Fellowship Initiative (PIFI), Grant No. 2024PVC0088. This project has received funding from the European Research Council (ERC) under the European Union’s Horizon 2020 research and innovation program (grant agreement No. [951549]). BC and MZ acknowledge the OPUS-LAP/GAČR-LA bilateral project (2021/43/I/ST9/01352/OPUS 22 and GF23-04053L).  MLM-A acknowledges financial support from Millenium Nucleus NCN19-058 and NCN2023${\_}$002 (TITANs) an ANID Millennium Science Initiative (AIM23-0001). SP acknowledges the financial support of the Conselho Nacional de Desenvolvimento Científico e Tecnológico (CNPq) Fellowship 301628/2024-6 and is supported by the International Gemini Observatory, a program of NSF NOIRLab, which is managed by the Association of Universities for Research in Astronomy (AURA) under a cooperative agreement with the U.S. National Science Foundation, on behalf of the Gemini partnership of Argentina, Brazil, Canada, Chile, the Republic of Korea, and the United States of America.. 32 spectroscopic observations of HE 0413-4031 used in this paper were obtained with the Southern African Large Telescope (SALT). They were made with the SALT under programs 2012-2-POL-003, 2013-1-POL-RSA-002, 2013-2-POL-RSA-001, 2014-1-POL-RSA-001, 2014-2-SCI-004, 2015-1-SCI-006, 2015-2-SCI-017, 2016-1-SCI-011, 2016-2-SCI-024, 2017-1-SCI-009, 2017-2-SCI-033, 2018-1-MLT-004 (PI: B. Czerny). Polish participation in SALT is funded by grant No. MEiN nr 2021/WK/01. 
\end{acknowledgments}

\vspace{5mm}

\software{CLOUDY \citep{cloudy23} }

\appendix
\section{Plots for the spectral fitting of RM 102 using power-law}\label{sect:pl_fit}
\begin{figure}
    \centering
    \includegraphics[width=18cm, height=10cm]{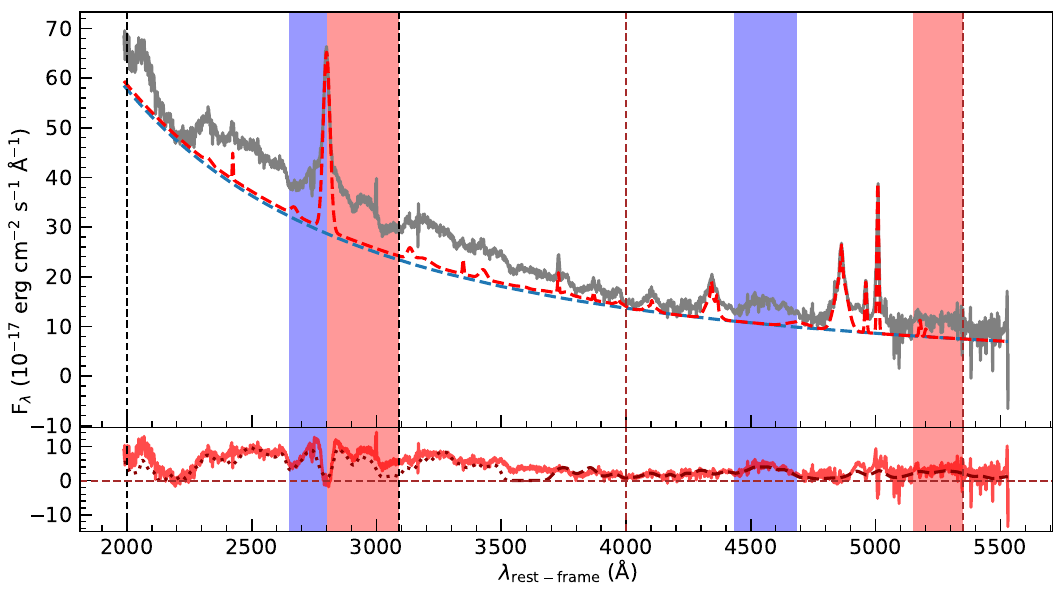}
    \caption{ \label{fig:MaryLoli} Observed composite spectrum of RM 102. The regions between the vertical black and brown dashed lines represent the broad UV and optical wavelength ranges, respectively, considered in this work. The shaded blue and red regions denote the blue and red parts in the corresponding regions. Top panel: the observed spectrum is in grey, the dashed red and the blue lines correspond to the best fit of the emission lines and the continuum, respectively. Bottom panel: the red line depicts the Fe II pseudocontinuum. For reference, we also show the semi-empirical UV (dotted line) and optical (dashed line) templates implemented in \texttt{PyQSOfit} in dark red.
   }
   
\end{figure}

\begin{figure*}
\centering
    \includegraphics[width=16cm, height=6cm]{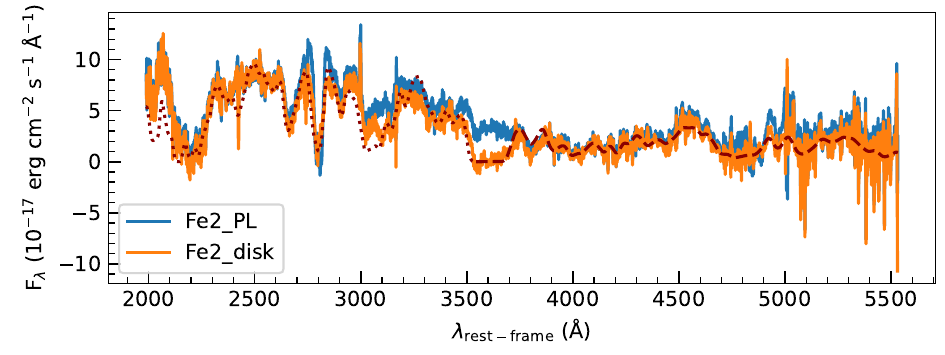}
    \caption{\label{fig:compare_PL_disk_fit}A comparison of the potential Fe II contribution from RM 102 using the power law (blue curve) and disk continuum (orange curve) fitting methods. For reference, we also plotted the semi-empirical UV (dotted line) and optical (dashed line) templates implemented in \texttt{PyQSOfit} in dark red. }
\end{figure*}

\section{EW contours for microturbulence velocities 20 and 50 km s$^{-1}$}\label{appendix:A}
\begin{figure*}
\centering
\includegraphics[width=18cm, height=16cm]{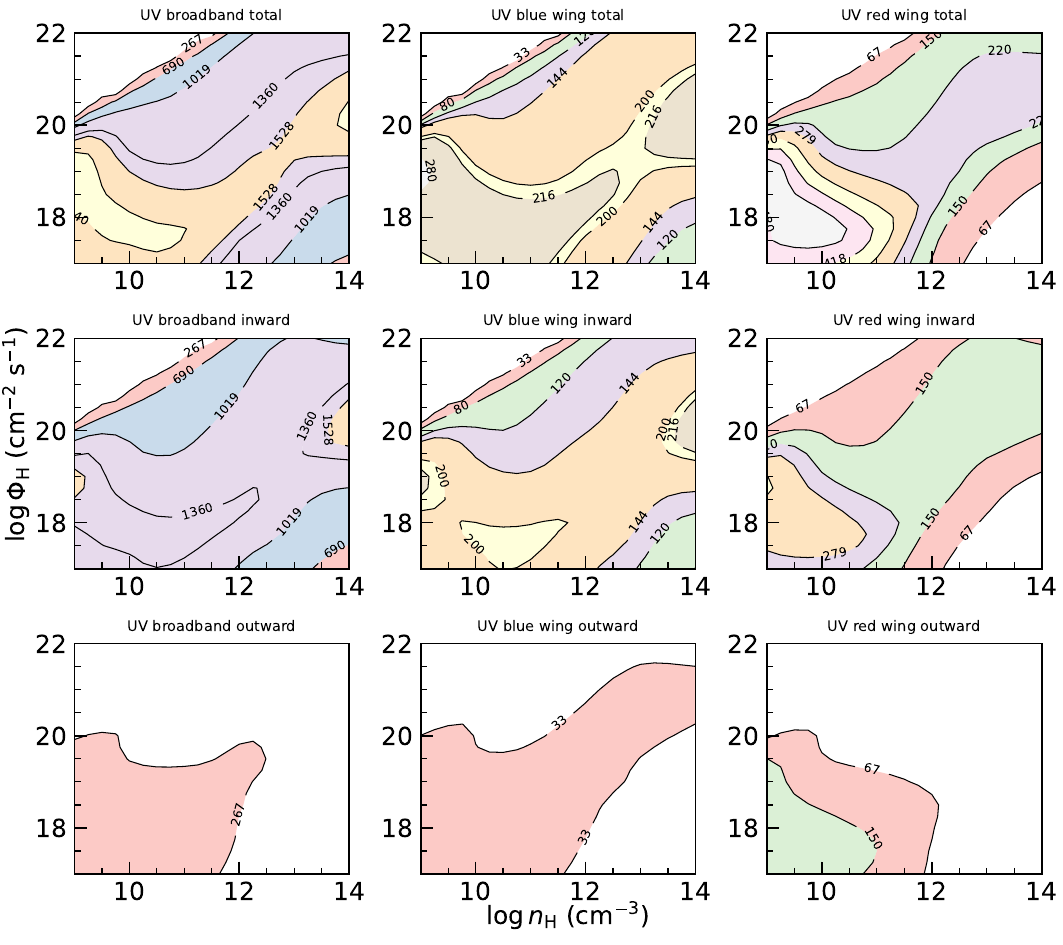}
\caption{\label{}Contour plots for the EW of UV Fe II blends for microturbulence V$_{\rm turb} = 20$ km/s.}
\end{figure*}

\begin{figure*}
\centering
\includegraphics[width=18cm, height=16cm]{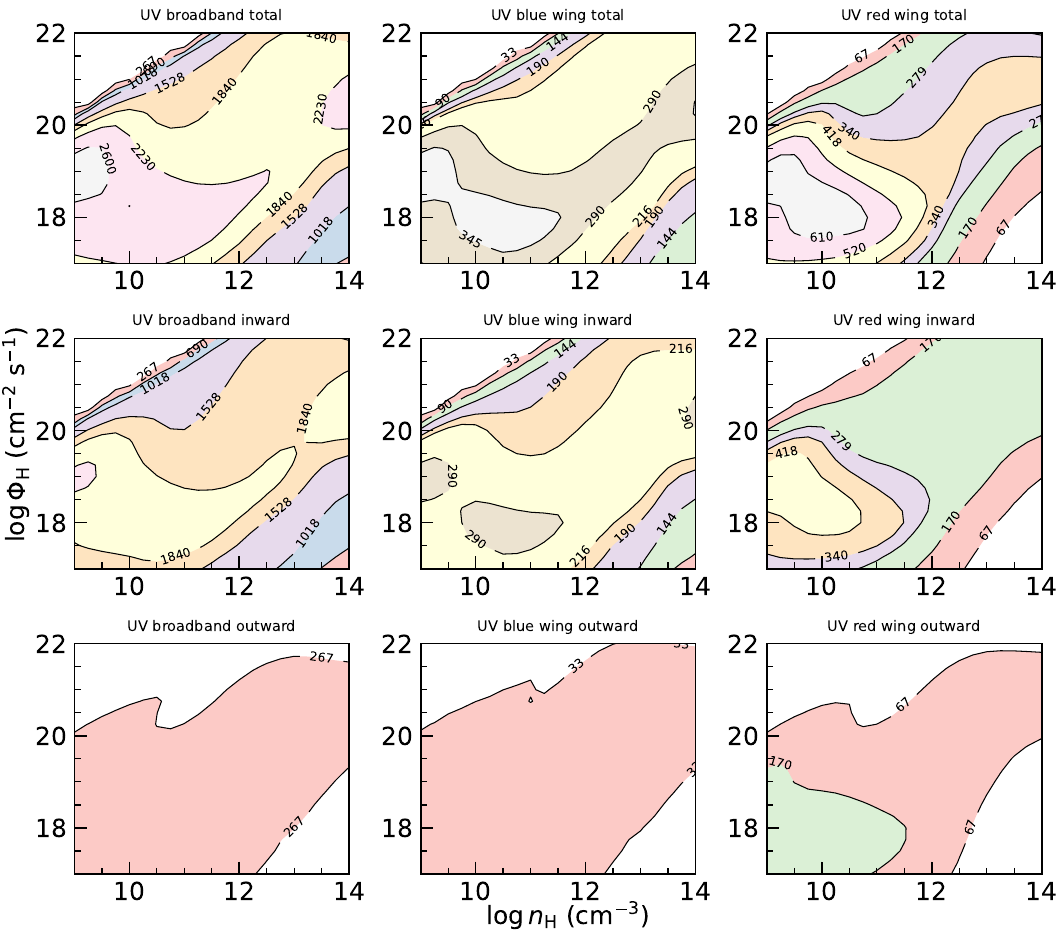}
\caption{\label{}Contour plots for the EW of UV Fe II blends for microturbulence V$_{\rm turb} = 50$ km/s.}
\end{figure*}

\begin{figure*}
\centering
\includegraphics[width=18cm, height=16cm]{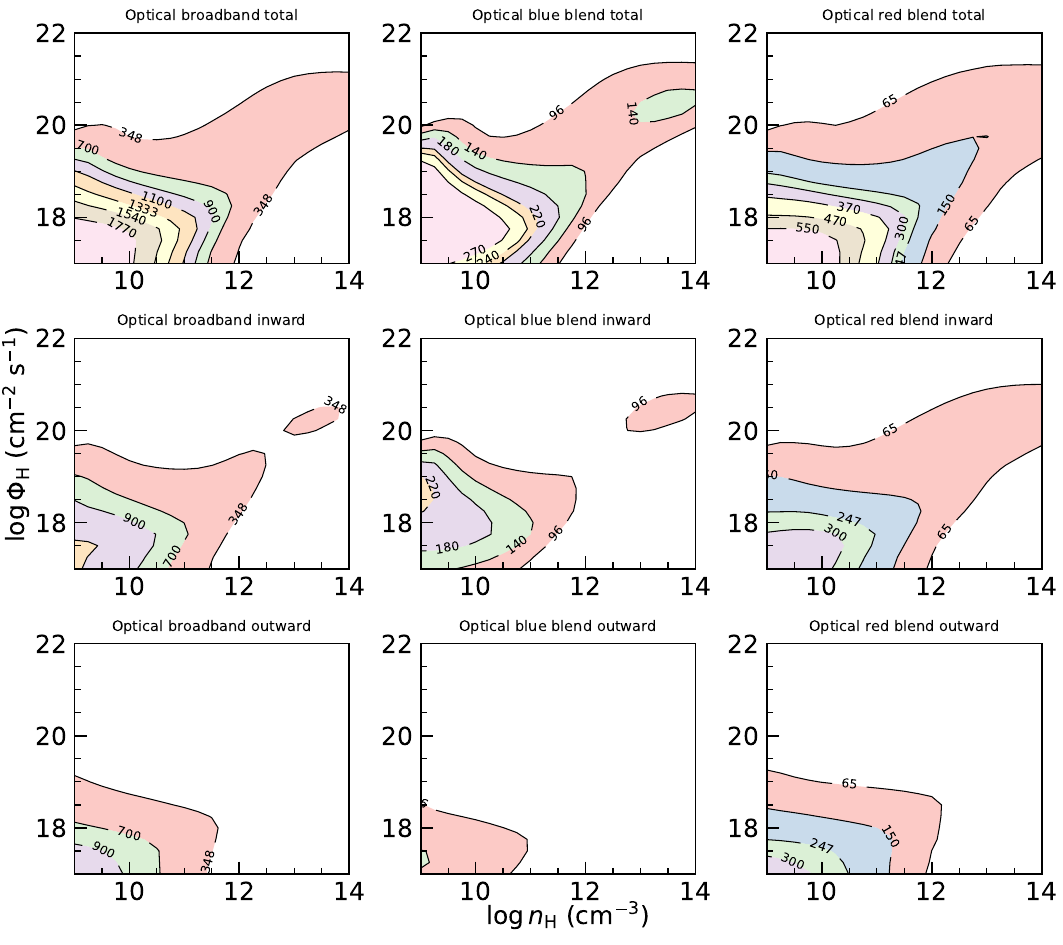}
\caption{\label{}Contour plots for the EW of Optical Fe II blends for microturbulence V$_{\rm turb} = 20$ km/s.}
\end{figure*}

\begin{figure*}
\centering
\includegraphics[width=18cm, height=16cm]{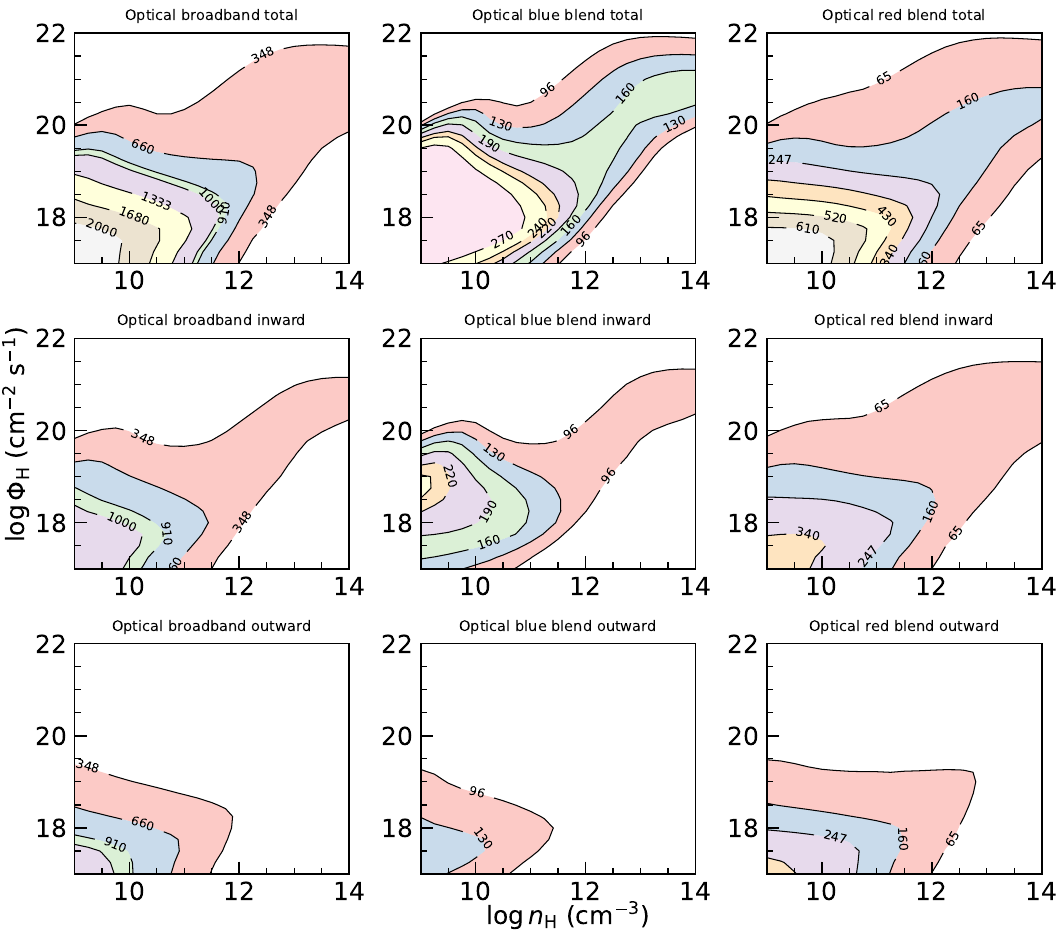}
\caption{\label{}Contour plots for the EW of Optical Fe II blends for microturbulence V$_{\rm turb} = 50$ km/s.}
\end{figure*}

\bibliography{master}{}
\bibliographystyle{aasjournal}


\end{document}